\documentclass[10pt,journal,compsoc]{IEEEtran}
	\usepackage{cite}
	\ifCLASSINFOpdf
	\else
	\fi
	\usepackage{enumitem}
	\usepackage{soul}
	
	\usepackage{cite}
	\usepackage{amsmath}
	\usepackage{soul,color}
	\usepackage{hyperref}

\usepackage{subfig}
	
	\usepackage{ragged2e}
	\usepackage{amsthm}
	\usepackage[T1]{fontenc}
	\usepackage{verbatim}
	\usepackage{prettyref}
	\usepackage{booktabs}
	\usepackage{textcomp}
	\usepackage{multirow}
	\usepackage{amstext}
	\usepackage{graphicx}
	\usepackage{amsmath}
	\usepackage{amsthm}
	\usepackage{amssymb}
	
	\usepackage[table]{xcolor}
	\usepackage{tikz}
	\usetikzlibrary{shapes,arrows}
	
	\tikzstyle{line}=[draw] 

	\usepackage{graphicx}
	\definecolor{lightgreen}{rgb}{0.56, 0.93, 0.56}
	\definecolor{mintgreen}{rgb}{0.6, 1.0, 0.6}
	
	\newtheorem{theorem}{Theorem}
	
	\usepackage{color}
	\usepackage[ruled,linesnumbered]{algorithm2e}
	\definecolor{seagreen}{rgb}{0.18, 0.55, 0.24}
	\SetAlFnt{\small\color{black}\tt}
	\LinesNumbered

\begin{document}
	%
	\title{Efficient and Privacy-Preserving Ridesharing Organization for Transferable
	and Non-Transferable Services}
	\author{Mahmoud~Nabil,~Ahmed~Sherif~\IEEEmembership{Member~IEEE}, ~Mohamed~Mahmoud~\IEEEmembership{Member~IEEE},~Ahmad~Alsharif~\IEEEmembership{Member~IEEE},~and\\~Mohamed~Abdallah~\IEEEmembership{Senior~Member,~IEEE,}
	\IEEEcompsocitemizethanks{\IEEEcompsocthanksitem Mahmoud Nabil ,  Mohamed Mahmud are with the Department of Electrical \& Computer
	Engineering, Tennessee Tech University, Cookeville, TN 38505 USA.\protect\\
	E-mails:\enskip{}\href{mailto:mnmahmoud42@students.tntech.edu}{mnmahmoud42@students.tntech.edu},
	\href{mailto:mmahmoud@tntech.edu}{mmahmoud@tntech.edu}.
	\IEEEcompsocthanksitem Ahmad Sherif is with the School of Computing Sciences and Computer Engineering,
	University of Southern Mississippi, Hattiesburg, 	MS, 39406 USA.
	E-mail:\enskip{}\href{mailto:ahmed.sherif@usm.edu}{ahmed.sherif@usm.edu},
	
	\IEEEcompsocthanksitem Ahmed Alsharif is with the Department of Computer Science, University of Central Arkansas, Conway, AR, 72035 USA and also with the Department of Electrical \& Computer Engineering,  Tennessee Tech University, Cookeville, TN 38505 USA.
	E-mail:\enskip{}\href{mailto:aalsharif@uca.edu}
	{aalsharif@uca.edu},

	\IEEEcompsocthanksitem Mohamed Abdallah is with the College of Science and Engineering
	Information \& Computing Technology at Hamad Bin Khalifa University, Doha, Qatar.
	E-mail:\enskip{}\href{mailto:moabdallah@qf.org.qa}
	{moabdallah@qf.org.qa},
	
	}
	
	\thanks{Manuscript received November xx, 2018; revised November xx, 2018.}

	}
	
	%
	%

	\markboth{IEEE TRANSACTIONS ON DEPENDAPLE AND SECURE COMPUTING,~Vol.~XX, No.~XX, November~2018}%
	{Shell \MakeLowercase{\textit{et al.}}: Bare Demo of IEEEtran.cls for Computer Society Journals}
	%



	\IEEEtitleabstractindextext{%
	\begin{abstract}
		Ridesharing allows multiple persons to share one vehicle for their trips instead of using multiple vehicles. Ridesharing can reduce the number of vehicles in the street, which consequently can reduce air pollution, traffic congestion, and transportation cost. However, ridesharing organization requires passengers to report sensitive location information about their trips to a trip organizing server (TOS) which creates a serious privacy issue.
		The existing ridesharing organization schemes are neither flexible nor scalable in the sense that they require a driver and a rider to have exactly the same trip to share a ride, and they are inefficient if applied to large geographic areas. In this paper, we propose two efficient privacy-preserving ridesharing organization schemes for Non-transferable Ridesharing Service (NRS) and Transferable Ridesharing Service (TRS).
		In NRS, a rider shares a ride from his/her trip's start to the destination with only one driver, whereas, in TRS, a rider can transfer between multiple drivers while en route until he reaches his destination.  In the proposed schemes, the ridesharing area is divided into a number of small geographic areas, called cells, and each cell has a unique identifier. Each driver/rider should encrypt his/her trip's data with modified kNN encryption scheme, and send an encrypted ridesharing offer/request to the TOS. In NRS scheme, Bloom filters are used to represent the trip information compactly before encryption. Then, the TOS can measure the similarity of the encrypted trips to organize shared rides without revealing either the users' identities or the locations. In TRS scheme, drivers report their encrypted routes, and then the TOS builds a directed graph that is passed to a modified version of Dijkstra's shortest path algorithm to search for an optimal path for rides that can achieve a set of preferences prescribed by the riders. Although TRS can be used to organize non-transferable trips, performance evaluation shows that NRS requires less communication overhead than TRS. Our formal privacy proof and analysis demonstrate that the proposed schemes can preserve users privacy and our experimental results using routes extracted from real maps show that the proposed schemes can be used efficiently for large cities.
	\end{abstract}
	
	
	}

	\maketitle

	\IEEEdisplaynontitleabstractindextext

	%
	\IEEEpeerreviewmaketitle

	\IEEEraisesectionheading{\section{Introduction}\label{sec:introduction}}

	%
	%
	%
	%
	\IEEEPARstart{R}{idesharing}, also known as carpooling,  is a service that enables multiple persons who have similar trips at the same time to travel together in one vehicle instead of using multiple vehicles.  It can reduce air pollution and traffic congestion by reducing the number of vehicles traveling on the roads. It can also reduce the trip cost by splitting the cost among several persons. Recently, the popularity of ridesharing has significantly increased \cite{rna1,ri3,ri6,agatz2012optimization,furuhata2013ridesharing,agatz2011dynamic,murray2004system,chan2012ridesharing}.
	As of 2010, there were at least 613 ridesharing organization platforms in North America mostly based on the internet \cite{xu2015traffic,deakin2010markets,furuhata2013ridesharing}.
	Furthermore, government policies have also been made in many countries to encourage citizens to share rides. Making High Occupancy Vehicle (HOV) lanes are one of such strategies. It allows the vehicles with more than two persons on-board to use a privileged lane \cite{hov10, hov11,hov12}.
	Another strategy to motivate sharing rides is providing toll discounts and reimbursements \cite{toll_discount}.

	Through smartphones, GPS systems, and Internet connectivity, organizing shared rides can be greatly facilitated. Users are required to register with an online platform that organizes shared rides, and then a user having a vehicle (driver) and seeking to share a ride with other users (riders) sends a ridesharing offer to a Trip Organizing Server (TOS). This offer should include the trip data, such as the starting location, destination, trip time and route. Also, riders looking for shared rides should first send ridesharing requests with similar information to the TOS. The TOS, then, matches the drivers' offers with riders' requests to assign one or multiple riders to each driver. Nevertheless, the TOS is run and operated by a private company that may be interested to collect information about the users' locations and activities.

	
	Different privacy-preserving ridesharing schemes are proposed in the literature \cite{csfHallgren,goel2017optimal,Aivodji2018,Bilogrevic2014,He2018,ri22,Wang2018}. However, existing schemes suffer from the following issues. First, some schemes \cite{ri22} and \cite{Wang2018} are not efficient when applied to a large geographic area because the trip data is very large due to representing each small area in the city, called cell, in the ridesharing request. Second, the schemes \cite{csfHallgren,goel2017optimal,Aivodji2018,Bilogrevic2014,He2018,ri22,Wang2018} are not flexible in the sense that they do not allow the riders to prescribe their ridesharing preferences, such as the trip length, transferable/non transferable services, the maximum number of transfers, etc. Finally, existing schemes \cite{csfHallgren,goel2017optimal,Aivodji2018,Bilogrevic2014,He2018,ri22,Wang2018} consider only non-transferable rides and cannot organize transferable rides. As will be demonstrated in the performance analysis, transferable rides can increase the number of served riders, and thus increase vehicle occupancy.
	
\begin{figure}[t]
		\centering
		\includegraphics[clip,scale=0.38]{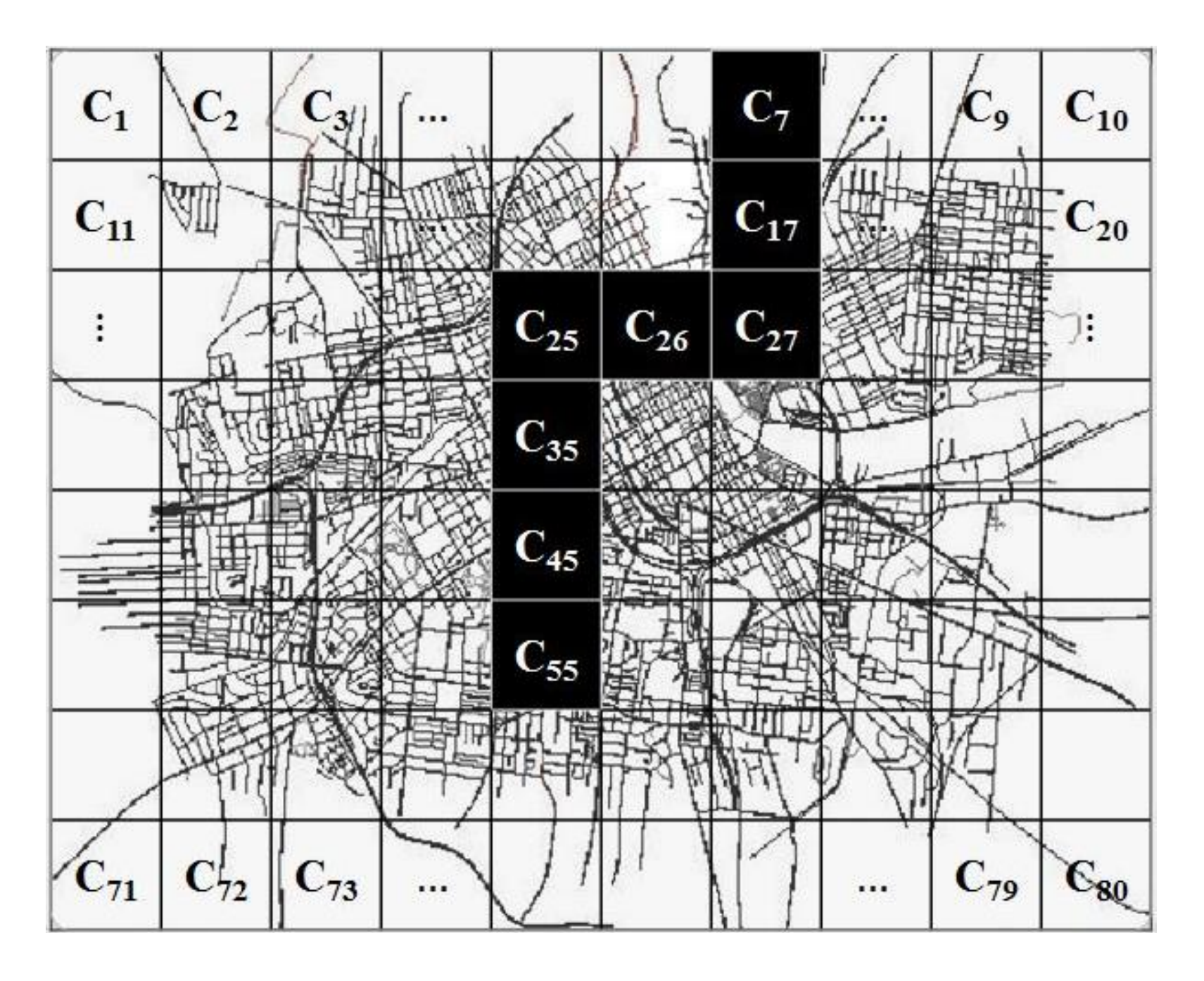}
		\vspace{-3mm}
		\caption{Dividing a ridesharing area into cells.}
		\label{fig:mapa} 
	\end{figure}

	In this paper, we aim to address the issues mentioned above by proposing two schemes for the organization of shared rides for non-transferable ridesharing service (NRS) and transferable ridesharing service (TRS). 
To the best of our knowledge, our TRS scheme is the first scheme that offers transferable ridesharing service. 
Both schemes enable the TOS to organize shared rides without knowing any private information about the users' locations and identities.
In the NRS scheme, a rider can share a ride with only one driver, while in TRS, the rider may need to take multiple vehicles to reach his destination, which can improve the ridesharing success rate. 
In the latter case, the rider has to transfer between multiple drivers during his trip. 
In our schemes, the users can select the type of service they need., e.g., NRS might be preferable to elderly and disabled people. 
Also, some passengers may prefer NRS because it is more comfortable and if they do not find a trip, they have backup plans by paying more, e.g., by taking a cab or driving their cars. 
Some passengers do not have such backup plans and they need the most comfortable TRS trip.
In TRS, users can also limit the trip length and the number of transfers of the trip that is returned by the TOS. In both schemes, the ridesharing area is divided into cells (geographic regions), and a unique identification number (ID) is assigned to each cell as shown in Figure \ref{fig:mapa}.
	
	For NRS organization, riders and drivers should submit to the TOS binary vectors representing their trip data encrypted by a modified kNN encryption scheme\cite{nabil}. The modified kNN encryption scheme ensures that each user in the system (driver/rider) has its own key including the TOS.
	Then, the TOS performs similarity measurement on the encrypted trip vectors to organize shared rides without learning any location information to preserve users' privacy.
	In addition, only the TOS can compute the similarty of any two ciphertexts. A Bloom filter \cite{bloom1970space} is used to compactly store the trip data to reduce the communication overhead and create the binary vector needed for the kNN similarity measurement.
	On the other hand, in TRS, drivers should report the encryption of the individual cells on their routes.
	Then, the TOS builds a directed graph using the drivers' individual cells.
	After that, a modified Dijkstra's searching algorithm  \cite{ri24} is used to search this graph to determine the route that can achieve the rider's preferences, constraints, and requirements without learning the identities or the locations of the users. For example, a rider can request a trip with the minimum number of cells regardless of the number of transfers.
	
	Compared to the existing schemes, our contributions can be summarized as follows.
	\begin{itemize}
		\item \textit{Transferable and non-transferable rides.}
		Two privacy-preserving ridesharing organization schemes are proposed TRS and NRS.  To the best of our knowledge, this is the first work that proposes the privacy-preserving ridesharing organization for the two cases. Although the focus of our schemes is on organizing shared rides (carpooling), the proposed schemes can also be adapted to work with ride-hailing services \cite{Pham2017} and \cite{oride2017}, such as Uber and Lyft.
	
		\item \textit{Efficiency and scalability}. Our schemes use secure and lightweight operations to enable TOS to efficiently find the riders and the drivers who can share rides in case of large cities. In addition, our schemes require low communication overhead.
		 \item \textit{Formal proof and analysis.}
		 We provide a formal proof and privacy analysis for the proposed schemes to demonstrate that NRS and TRS can organize shared rides without disclosing locations and identities information.
	
		 \item \textit{Simulation with real maps.}
		 Proposed schemes were implemented with MATLAB using real map and routes. The experimental results demonstrate that the communication and storage overheads are acceptable, and our schemes are scalable and can be used efficiently in cases of large cities. In addition,  Although TRS can be used to organize both transferable and non-transferable rides, The NRS provides better efficiency in terms of the communication overhead for non-transferable services.
	\end{itemize}{}


	The remainder of this paper is organized as follows. The network and threat models are introduced in \prettyref{sec:system_model}.
	Preliminaries and design goals are given in \prettyref{sec:prel}.
	The proposed schemes are presented in \prettyref{sec:Proposed-Scheme}. Privacy analysis is discussed in \prettyref{sec:privacy_analysis}.
	Performance evaluations are given in \prettyref{sec:performance}.
	The related works are presented in \prettyref{sec:related Work}.
	Finally, conclusions are drawn in \prettyref{sec:conclusion}.

	\section{System Model and Design Goals} \label{sec:system_model}
	
	\subsection{Network Model}

\begin{figure}[!t]
		\includegraphics[width=0.95\columnwidth,height=0.38\textwidth]{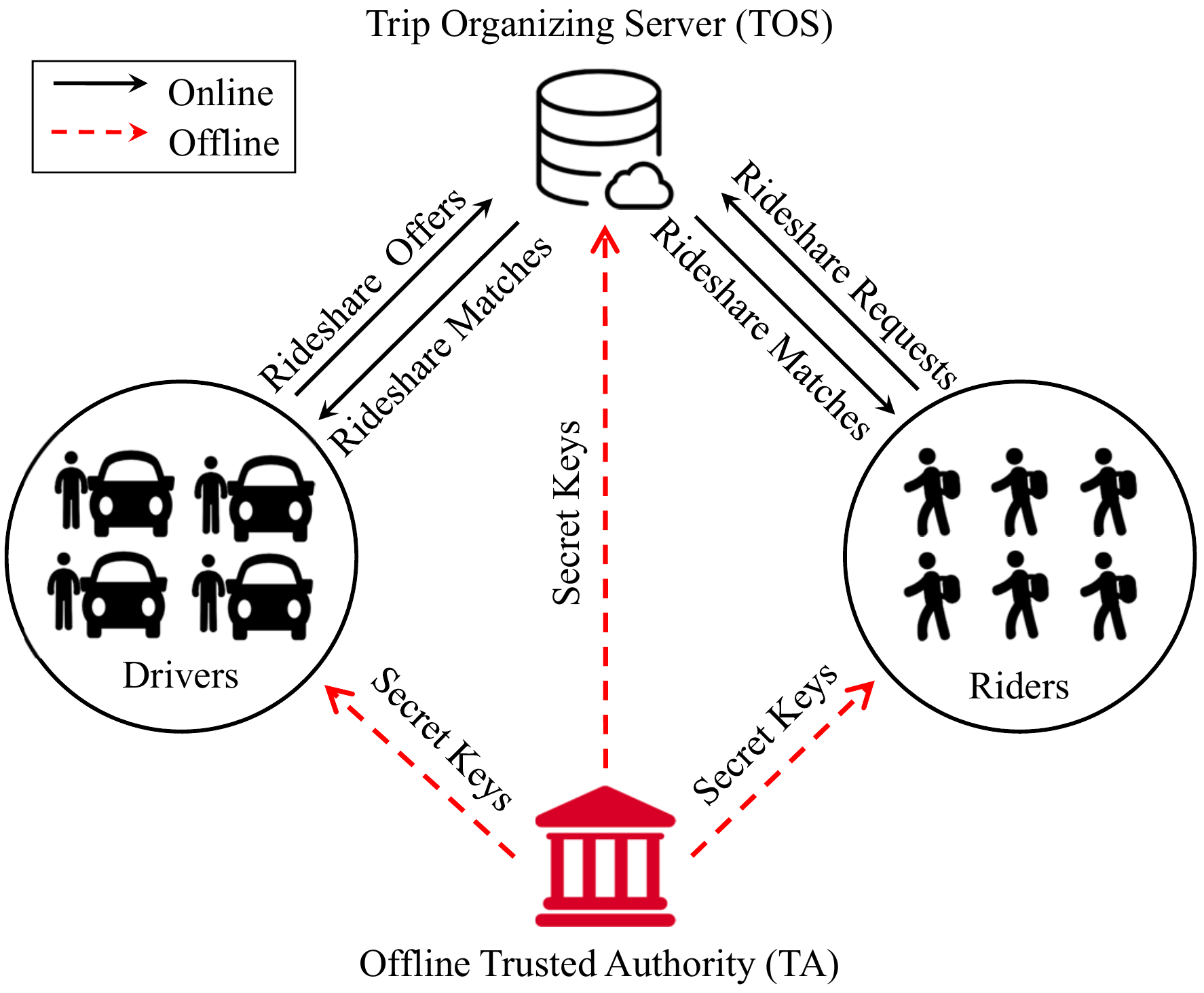}
		\vspace{-2mm}
		\caption{Considered network model.}
		\label{fig:network_model} \vspace{-5mm}
	\end{figure}

	As shown in \autoref{fig:network_model}, the considered network model consists of an offline Trusted Authority (TA), drivers, riders and a TOS.
	The TA is responsible for generating and distributing unique secret keys to drivers, riders, and the TOS. Drivers and riders can use a smartphone application to send their encrypted trip data to the TOS through the internet. The TOS should execute one of the proposed schemes based on the drivers 'and riders' preferences and connects the drivers and riders that can share rides. \autoref{fig:mapa} shows the ridesharing area considered in our simulation. 
To achieve forward and backward privacy as will be discussed in \autoref{sec:privacy_analysis}, the cell's identifiers should frequently change (e.g., every day) so that the TOS cannot link old trips' data to the new data to obtain side information.

	\subsection{Threat Model}
	
	\label{Threat Model}
	In our schemes, we consider the trusted authority is trustworthy and cannot be compromised. 
Additionally, we consider that drivers, riders, and the TOS are "honest-but-curious", i.e., they follow the proposed schemes and do not aim to disrupt their operation, however, they are curious to learn useful information about users, such as the trip's pick-up/drop-off locations and times. 
The attackers are also interested in learning the identities of users who share rides with each other. 
In addition, we assume that the attackers can be internal or external. 
The TOS and legitimate users can be internal attackers. 
External attackers may try to pretend to be a driver/rider and send offers/requests to the TOS to expose the privacy of other riders/drivers. 
They can also launch attacks like eavesdropping and impersonation, etc., to learn sensitive information about the users' locations, routes, and identities. 
In our schemes, TOS is a third party known to all riders, drivers, and TA.

	\subsection{Design Goals}
	
	\label{design_goals} Our goal is to develop two privacy-preserving ridesharing organization schemes for NRS and TRS with the following objectives.

	\begin{itemize}
	
	\item \emph{Transferable and non-transferable services}:
     Non-transferable service may be preferred by some users, such as elderly, while, transferable service can increase the chance of organizing shared rides.
	
	\item \textit{Efficiency and scalability.}
	The schemes should be efficient in case of a large ridesharing region (i.e. big cities) and should process large number of ridesharing requests quickly.
	
	\item \textit{User requirements and preferences.}
	Users should be able to select the type of service (NRS or TRS) and prescribe their search preference such as trip length, maximum number of transfers, etc.
	
	\item \textit{Trip data privacy.}
	The TOS should be able to organize shared rides without learning the pick-up/drop-off locations and times or even any road segment in users trips.
	Eavesdroppers and malicious users should not be able to learn any location/time information about
	other users' trips.
	
	\item \textit{Forward and backward privacy}:
	To prevent the TOS from inferring incremental side information over time, by matching the trip's data sent at different times, the encrypted trip data sent in a given day cannot be matched to old or future trips' data.
	
	\item \textit{Data linkability:}
	(1) \textit{Requests-user un-linkability}: Given different ridesharing requests sent from a user, the TOS should not learn if these requests are sent from the same user or not;
	(2) \textit{Driver-rider pair un-linkability}:
	If a driver-rider pair shares a ride, the TOS should not identify the same pair when they share rides in the future;
	(3)\textit{ Same requests un-linkability}:
	The TOS should not link the ridesharing requests (or offers) of the same trip at different
	times.
	\item \textit{Users' anonimity}:
	The TOS should ensure that the received requests/offers are coming from legitimate users, but, it should not be able to reveal the real identity of the users from the credentials they use for authentication or the locations they visit.

	\end{itemize}
	\begin{table}[!t]
	\centering
	\caption{List of notations and their meanings.\label{table:List-of-Notations}}
	\begin{tabular}{ll}
	\toprule
	Notation & Description \\
	\hline\\
	\scalebox{0.75}{$K_{NT}=\left\{M_{1},M_{2},N_{1}, \dots, N_{8}\right\}$} & \scalebox{0.9}{TA Master Secret Matrices for NRS scheme}\\[0.3em]
	\scalebox{0.75}{$K_{T}=\left\{V_{1},V_{2},T_{1}, \dots, T_{8}\right\}$} & \scalebox{0.9}{TA Master Secret Matrices for TRS scheme}\\[0.3em]
	\scalebox{0.95}{$S_{NT},S_{T}$}   & \scalebox{0.9}{Splitting Indicator Vector for NRS and TRS}\\[0.3em]
	\scalebox{0.95}{$X,Y$}            & \scalebox{0.9}{TOS secrets for NRS}\\[0.3em]
    	\scalebox{0.95}{$W,Z$}            & \scalebox{0.9}{TOS secrets for TRS}\\[0.3em]
	\scalebox{0.9}{$\mathcal{DNSK}$} & \scalebox{0.9}{Driver Secret key in NRS}\\[0.3em]
	\scalebox{0.9}{$\mathcal{RNSK}$} & \scalebox{0.9}{Rider Secret key in NRS}\\[0.3em]
	\scalebox{0.9}{$\mathcal{DTSK}$} & \scalebox{0.9}{Driver Secret key in TRS}\\[0.3em]
	\scalebox{0.9}{$\mathcal{RTSK}$} & \scalebox{0.9}{Rider Secret key in TRS}\\[0.3em]
	\scalebox{0.95}{$m$} 	  & \scalebox{0.9}{Number of bits needed for the Bloom filter}\\[0.3em]
	\scalebox{0.95}{$k$} 	  & \scalebox{0.84}{Number of bits needed to represent an TRS cell}\\[0.3em]
	\scalebox{0.95}{$\ensuremath{\ell}$} 	  & \scalebox{0.9}{Number of bits needed to represent time}\\[0.3em]
	\scalebox{0.95}{$n$} 	  & \scalebox{0.9}{Number of bits needed for an TRS vector}\\[0.3em]
    \scalebox{0.95}{$N_r$} 	  & \scalebox{0.9}{Number of riders with each driver in NRS}\\[0.3em]
    \scalebox{0.95}{$n\_d$} 	  & \scalebox{0.9}{Number of drivers}\\[0.3em]
    \scalebox{0.95}{$k\_r$} 	  & \scalebox{0.9}{Number of riders}\\[0.3em]
	\scalebox{0.95}{$C_p, C_d, r, t$} &
	\scalebox{0.95}{Pick-up cell, drop-off cell, route, time}\\[0.3em]
	\scalebox{0.95}{$I_{D}^{(C_p)},I_{D}^{(C_d)},I_{D}^{(r)},I_{D}^{(t)}$} & \scalebox{0.82}{Driver D: pick-up, drop-off, route and time indices}\\[0.3em]
	\scalebox{0.95}{$I_{R}^{(C_p)},I_{R}^{(C_d)},I_{R}^{(r)},I_{R}^{(t)}$} & \scalebox{0.83}{Rider R: pick-up, drop-off, route and time indices}\\[0.3em]
	\bottomrule
	\end{tabular}
	\end{table}

	\section{Proposed Schemes\label{sec:Proposed-Scheme}}
	This section presents the details of the proposed schemes.
	For better readability, the main notations used throughout this section are given in Table \ref{table:List-of-Notations}
	\subsection{System Bootstrap \label{subsec:System-Bootstrap}}
	The TA chooses two sets of keys as master secret keys,
	$K_{NT}$ for NRS and $K_{T}$ for TRS, where
	$K_{NT}=\left\{M_{1},M_{2},N_{1}, \dots, N_{8}\right\}$
	and
	$K_{T}=\left\{V_{1},V_{2},T_{1}, \dots, T_{8}\right\}$.
	$S_{NT}$ and $S_T$ are binary vectors used as splitting indicators during the encryption process, as will be explained later in this section.
	The size of $S_{NT}$ is $m$, where $m$ is the size of the binary vector to be encrypted in NRS scheme.
	All other elements in the set $K_{NT}$ are $m\times m$ invertible matrices.
	Similarly, the size of $S_T$ is $n$, where $n$ is the size of the binary vector to be encrypted in TRS scheme and all other elements in the set, $K_T$, are $n\times n$ invertible matrices of random secret numbers.

	
	Then, the TA generates the TOS' secret keys. The keys consist of four random matrices $X, Y, W$, and $Z$.
	The two matrices $X$ and $Y$ are to be used in NRS while $W$ and $Z$ are to be used in TRS. Note that,
	$X$ and $Y$ are of size $m \times m$, while $W$ and $Z$ are of size $n \times n$.

	Each user in the system can play the role of a driver or a rider. Also, each user can offer/request non-transferable or transferable service. Using the master secret sets and the TOS secrets, the TA computes \textit{unique} secret keys for each user in the system as follows.

	\begin{enumerate}[leftmargin=5mm]
		
		\item \textbf{D}river in \textbf{N}RS \textbf{S}ecret \textbf{K}ey:
		\begin{equation*}\hspace*{-3mm}
		\begin {split}
		\mathcal{DNSK}=\{
		&Y^{\text{-1}}N_{1}^{\text{-1}}A_{N},\ Y^{\text{-1}}N_{2}^{\text{-1}}B_{N},\ Y^{\text{-1}}N_{3}^{\text{-1}}A_{N},\\
		&Y^{\text{-1}}N_{4}^{\text{-1}}B_{N},\ Y^{\text{-1}}N_{5}^{\text{-1}}C_{N},\ Y^{\text{-1}}N_{6}^{\text{-1}}D_{N},\\
		&Y^{\text{-1}}N_{7}^{\text{-1}}C_{N},\ Y^{\text{-1}}N_{8}^{\text{-1}}D_{N}\}
		\end{split}
		\end{equation*}
		where $A_{N},B_{N}$, $C_{N}$, and $D_{N}$ are random and invertible matrices of size $m \times m$, such that $A_{N}+B_{N}=M_{1}^{\text{-1}}$, and  $C_{N}+D_{N}=M_{2}^{\text{-1}}$. Note that, $\mathcal{DNSK}$ key set has eight elements where each element is a matrix of size $m \times m$.
		
		\item \textbf{R}ider in \textbf{N}RS \textbf{S}ecret \textbf{K}ey:
		\begin{equation*}\hspace*{-2mm}
		\begin {split}
		\mathcal{RNSK}=\{
		&E_{N}N_{1}X,\ E_{N}N_{2}X,\ F_{N}N_{3}X,\ F_{N}N_{4}X,\\
		&G_{N}N_{5}X,\ G_{N}N_{6}X,\ H_{N}N_{7}X,\ H_{N}N_{8}X\}
		\end{split}
		\end{equation*}
		
		where $E_{N},F_{N},G_{N}$, and $H_{N}$ are random and invertible matrices of size $m \times m$, such that $E_{N}+F_{N}=M_{1}$, and  $G_{N}+H_{N}=M_{2}$. Note that, $\mathcal{RNSK}$ key set has eight elements where each element is a matrix of size $m \times m$.
		
		\item \textbf{D}river in \textbf{T}RS \textbf{S}ecret \textbf{K}ey:
		\begin{equation*}\hspace*{-3mm}
		\begin {split}
		\mathcal{DTSK}=\{
		&Z^{\text{-1}}T_{1}^{\text{-1}}A_{T},\ Z^{\text{-1}}T_{2}^{\text{-1}}B_{T},\ Z^{\text{-1}}T_{3}^{\text{-1}}A_{T},\\
		&Z^{\text{-1}}T_{4}^{\text{-1}}B_{T},\ Z^{\text{-1}}T_{5}^{\text{-1}}C_{T},\ Z^{\text{-1}}T_{6}^{\text{-1}}D_{T},\\
		&Z^{\text{-1}}T_{7}^{\text{-1}}C_{T},\ Z^{\text{-1}}T_{8}^{\text{-1}}D_{T}\}
		\end{split}
		\end{equation*}
		where $A_{T},B_{T},C_{T}$, and $D_{T}$ are random and invertible matrices of size $n \times n$, such that $A_{T}+B_{T}=V_{1}^{\text{-1}}$, and  $C_{T}+D_{T}=V_{2}^{\text{-1}}$.  Note that, $\mathcal{DTSK}$ key set has eight elements where each element is a matrix of size $n \times n$.
		
		\item \textbf{R}ider in \textbf{T}RS mode \textbf{S}ecret \textbf{K}ey:
		\begin{equation*}\hspace*{-3mm}
		\begin {split}
		\mathcal{RTSK}=\{
		&E_{T}T_{1}W,\ E_{T}T_{2}W,\ F_{T}T_{3}W,\ F_{T}T_{4}W,\\
		&G_{T}T_{5}W,\ G_{T}T_{6}W,\ H_{T}T_{7}W,\ H_{T}T_{8}W\}
		\end{split}
		\end{equation*}
		
		where $E_{T},F_{T},G_{T}$, and $H_{T}$ are random and invertible matrices of size $n \times n$, such that $E_{T}+F_{T}=V_{1}$, and  $G_{T}+H_{T}=V_{2}$.   Note that, $\mathcal{RTSK}$ key set has eight elements where each element is a matrix of size $n \times n$.
		
	\end{enumerate}
	In addition to the four key sets, each user receives from the TA a set of certified pseudonymous that are used for authentication with the TOS. In addition, each user should receive the two vectors $S_{NT}$ and $S_T$.
	
	%
	
	\subsection{Non-transferable Ride Sharing Organization}
	%

	\label{sec:Non-Transferable Ride Sharing Organization Scheme}
	
	\subsubsection{Overview} \label{Overview1}
The original version of kNN encryption scheme was proposed in \cite{ri21}, and we modified it in \cite{nabil}.
In the modified version, we can measure the similarity of the indices submitted from the riders with indices submitted from different drivers, while the original version can only measure the similarity between the riders' indices and one driver's indices.
The modified kNN technique is used for matching the riders' requests indices with drivers' offers indices to organize the shared rides.
In the proposed scheme, each driver creates an encrypted ridesharing offer, and each rider creates an encrypted ridesharing request.
	The TOS receives these offers and requests and computes their similarity to organize non-transferable shared rides without learning any location information.
	Finally, the TOS connects the drivers and riders that can share rides.
	
	
	\begin{figure}[!t]
	\center \includegraphics[clip,width=0.5\textwidth]{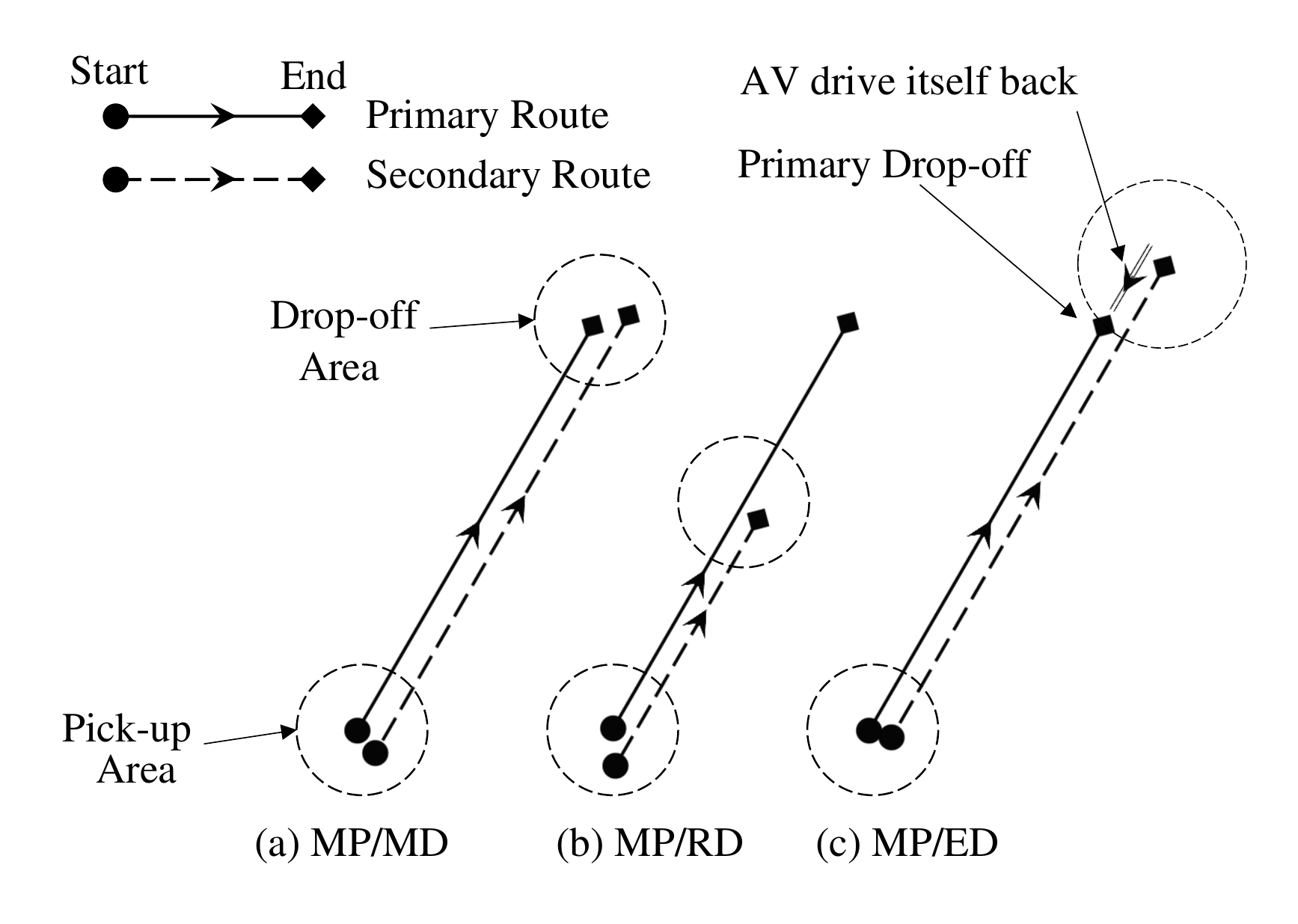}\vspace{-5mm}
	\caption{Ridesharing cases in NRS scheme.}
	\label{fig:cases} \vspace{-3mm}
	
	\end{figure}

	\subsubsection{Ride-Sharing Cases} \label{Ride-Sharing Cases}
	Three different ridesharing cases are considered for the NRS scheme as shown in \autoref{fig:cases}.
	
	\begin{itemize}[leftmargin=5mm]
		
		\item \textit{Matched Pick-up and Matched Drop-off} (MP/MD).
The driver decides the pick-up and drop-off areas of the riders.
These areas are in the surrounding of the driver's start and end locations. 		
As shown in \autoref{fig:cases}(a), the destination of the rider lies within the drop-off area defined by the driver.
		\vspace{2mm}
		\item \textit{Matched Pick-up and on-Route Drop-off} (MP/RD).
The driver decides the pick-up area like MP/MD case.
For the drop off, the driver can share a ride with a rider whose drop-off location is different from his trip's end location, but it lies on the surrounding of his route.
As shown in \autoref{fig:cases}(b), the destination of the rider lies on the driver's route.
In this case, the vehicle diverts from its route to drop off the rider(s) and then it drives back to the driver's route.
		\vspace{2mm}
	
		\item \textit{Matched Pick-up and Extended Drop-off} (MP/ED).
The driver decides the pick-up area like MP/MD and MP/RD cases.
In this case, the drop-off location of the rider does not lie on the surrounding area of his driver's route or end location.
As shown in \autoref{fig:cases}(c), the destination of the driver lies on the rider's route.
Therefore, the vehicle first drops the driver off, then it delivers the rider to his destination, and finally, it drives itself back to the driver's location.
This case is not possible in human-driven cars unless there is a human driver who can drive the car back to the driver after delivering the rider.
	\end{itemize}
	
	Note that, these ridesharing cases consider that the riders and drivers have the same pick-up area but they can be easily extended so that the driver can pick-up riders at any point on his route.

	
	
	\subsubsection{Driver's Offer Submission\label{subsec:Primary-User's-Offer} }
	

	Each driver should represent his pick-up area, drop-off area, and his route by a list of cell IDs.
	For example, Figure \ref{fig:mapa} shows a user's route that can be represented by the following set of cell IDs $\{C_{7},C_{17},C_{27},C_{26},C_{25},C_{35},C_{45},C_{55}\}$.
	
	The driver creates a ridesharing offer that has four encrypted vectors referred to as indices.
	These indices contain driver's trip data including  pick-up area index $I_{D}^{(C_p)}$,
	drop-off area index $I_{D}^{(C_d)}$, trip route index $I_{D}^{(r)}$, and pick-up time index $I_{D}^{(t)}$.

\begin{figure}[!t]
		\centering \scalebox{0.72}{
			
			\tikzstyle{block} = [rectangle, draw, fill=white,     text width=12em, text centered, rounded corners, minimum height=4em, line width=0.5mm] \tikzstyle{line} = [draw, line width=0.6mm, -latex']
			\tikzset{   font={\fontsize{11pt}{12}\selectfont}}
			
			\begin{tikzpicture}[node distance = 2cm, auto]
			\node [] (Begin) {$\{C_7,C17,\dots,C_{55}\}$};
			\node [left of=Begin, node distance=3cm] (name) {Route Cells:};
			\node [block,below of=Begin] (Build) {Bloom Filter Building Algorithm};
			\node [block, below of=Build,node distance=2.5cm] (Binary) {Splitting};
			\node [block, below of=Binary,node distance=2.5cm] (Encrypt) {Encryption};
			\node [below of=Encrypt] (Output) {$I_{D}^{(r)}$};
			\node [right of=Binary, node distance=4cm] (Key_S) {$S_{NT}$};
			\node [right of=Encrypt, node distance=4cm] (Key) {$\mathcal{DNSK}$};
	
			\path [line] (Begin) -- (Build) ;
			\path [line] (Build) -- (Binary) node[midway] {$P^{(r)}$};
			\path [line] (Binary) -- (Encrypt) node[midway] {$p'^{(r)},p''^{(r)}$};;
			\path [line] (Encrypt) -- (Output);
			\path [line] (Key) -- (Encrypt);
			\path [line] (Key_S) -- (Binary);
	
			\end{tikzpicture}
			
		}
		\vspace{-3mm}
		\caption{Route Data Encryption Process.}
		\label{fig:fig131}
	\end{figure}
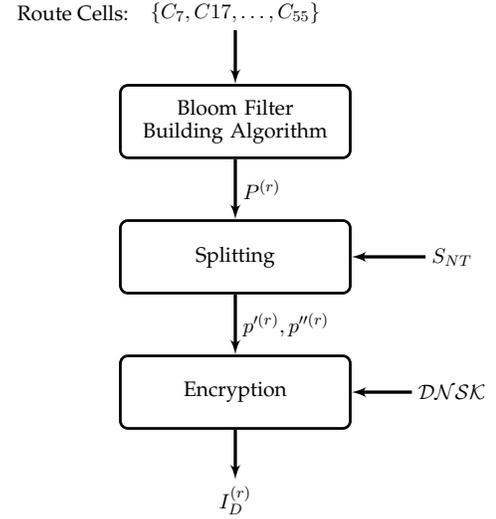

	\autoref{fig:fig131} shows the process of computing the trip route index $I_{D}^{(r)}$ by the driver.
	First, the route cells are passed to a Bloom filter building algorithm to generate a binary
	column vector $P^{(r)}$ of size $m$. Note that, the algorithm is modified to guarantee that the $\alpha$ hash values of each cell ID are distinct. For example, if $H_{i}(C)=H_{j}(C)$, a counter is concatenated to $C$  such that $\ H_{i}(C,counter)\neq H_{j}(C,counter)$, where $counter$ is the first number that can make the two hash values different \cite{bloom1970space}.
	Then, the driver uses the key set $\mathcal{DNSK}$, defined in \autoref{subsec:System-Bootstrap}, to encrypt $P^{(r)}$ using our modified kNN encryption scheme \cite{nabil}.
	For encryption, the driver uses $S_{NT}$ to split $P^{(r)}$  into two random column vectors $p'^{(r)}$ and $p''^{(r)}$ of the same size as $P^{(r)}$. The splitting is done as follows. If the $j^{th}$ bit of $S_{NT}$ is zero, then,  $p'^{(r)}(j)$ and $p''^{(r)}(j)$ are set similar to $P^{(r)}(j)$, while if it is one, $p'^{(r)}(j)$ and $p''^{(r)}(j)$ are set to two random numbers
	such that their summation is equal to $P^{(r)}(j)$.
	After splitting $P^{(r)}$, the encrypted index $I_{D}^{(r)}$ can be computed using the secret key $\mathcal{DNSK}$ and the vectors $p'^{(r)},p''^{(r)}$ as follows
	\begin{equation}\label{eq:ID}\hspace*{-3mm}
	\begin {split}
		I_{D}^{(r)}=\Big[
		&Y^{\text{-1}}N_{1}^{\text{-1}}A_{N}p'^{(r)};\ Y^{\text{-1}}N_{2}^{\text{-1}}B_{N}p'^{(r)};\ Y^{\text{-1}}N_{3}^{\text{-1}}A_{N}p'^{(r)};\\
		&Y^{\text{-1}}N_{4}^{\text{-1}}B_{N}p'^{(r)};\ Y^{\text{-1}}N_{5}^{\text{-1}}C_{N}p''^{(r)};\ Y^{\text{-1}}N_{6}^{\text{-1}}D_{N}p''^{(r)};\\
		&Y^{\text{-1}}N_{7}^{\text{-1}}C_{N}p''^{(r)};\ Y^{\text{-1}}N_{8}^{\text{-1}}D_{N}p''^{(r)}\Big]
	\end{split}
	\end{equation}
	where $I_{D}^{(r)}$ is a column vector of size $8$ elements, and each element is a column vector of size $m$.
	Note that, every time the same vector is encrypted, the ciphertext looks different due to the random numbers used in creating $p'^{(r)}$ and $p''^{(r)}$.

	Using the same process, the pick-up and drop-off indices, $I_{D}^{(C_p)}$ and $I_{D}^{(C_d)}$, can be computed by passing the pick-up cells and drop-off cells to the Bloom filter algorithm, and encrypting the Bloom filter vector.
	To generate the pick-up time index $I_{D}^{(t)}$, we use a binary vector where each bit represents one-time slot in the day such that all the bit values are zeros except the bit corresponding to the trip start time that should store one.
	Then, this binary vector is encrypted in the same manner to obtain $I_{D}^{(t)}$.

	Along with the indices, the driver should send to the TOS the maximum number of riders with whom he can share rides. Also, the driver should encrypt his contact information and send it to the TOS along an anonymous signature.
	Moreover, the driver can select some of the preferred ridesharing cases.
	For example, he might prefer only MP/RD case so that riders can not know his final destination, he might prefer only MP/MD case so that he never stops during the trip, or he can select more than one ridesharing case.

	\subsubsection{Rider's Request Submission\label{Secondary Users' Requests} }
	
	Each rider should generate a ridesharing request that has the pick-up location index $I_{R}^{(C_p)}$, drop-off location index $I_{R}^{(C_d)}$, trip route index $I_{R}^{(r)}$, and pick-up time index $I_{R}^{(t)}$.
	More specifically, to compute $I_{R}^{(C_p)}$, the rider's pick-up cell ID is passed to the Bloom filter building algorithm to generate a binary row vector $Q^{(C_p)}$.
	Since the rider's pick-up location is one cell, the Bloom filter vector contains exactly $\alpha$ bits set to 1.
	Then, $S_{NT}$ is used to split $Q^{(C_p)}$ into two random row vectors $q'^{(C_p)}$ and $q''^{(C_p)}$ of the same size as $Q^{(C_p)}$.
	The splitting process is opposite to that of the drivers.
	If the $j^{th}$ bit of $S_{NT}$ is one, then,  $q'^{(C_p)}(j)$ and $q''^{(C_p)}(j)$ are set similar to $Q^{(C_p)}(j)$, while if it is zero, $q'^{(C_p)}(j)$ and $q''^{(C_p)}(j)$ are set to two random numbers
	such that their summation is equal to $Q^{(C_p)}(j)$.
	Then, the rider can compute the encrypted index $I_{R}^{(C_p)}$ using the key set $\mathcal{RNSK}$ and the vectors $q'^{(C_p)}$ and $q''^{(C_p)}$ as follows.
	\begin{equation} \label{eq:IR}
	\begin {split}
	I_{R}^{(C_p)}=\Big[
	&q'^{(C_p)}E_{N}N_{1}X,\ q'^{(C_p)}E_{N}N_{2}X,\ q'^{(C_p)}F_{N}N_{3}X,\\
	&q'^{(C_p)}F_{N}N_{4}X,\ q''^{(C_p)}G_{N}N_{5}X,\ q''^{(C_p)}G_{N}N_{6}X,\\
	&q''^{(C_p)}H_{N}N_{7}X,\ q''^{(C_p)}H_{N}N_{8}X\Big]
	\end {split}
	\end{equation}
	where $I_{R}^{(C_p)}$ is a row vector with size $8$ elements, and each element is a row vector of size $m$.
	Using of a similar process, the rider can generate $I_{R}^{(t)}$, $I_{R}^{(C_d)}$,
	and $I_{R}^{(r)}$, for the pick-up time, drop-off location, and route respectively.
	Also, the rider should encrypt his contact information and send it to the TOS along with an anonymous signature.
	
	\subsubsection{Organizing Shared Rides \label{Organizing Shared Rides} }
	
	First, the TOS must ensure that the received offers and requests are coming from legitimate users.
	This can be achieved by verifying the received signatures associated with each offer/request.
	In addition, the TOS multiplies the elements of the offers' indices by $Y$ and elements of the requests' indices by $X^{\text{-1}}$ to remove its secrets $Y^{\text{-1}}$ and $X$ from the offers and requests respectively. Thus, only the TOS can match drivers' and riders' indices because its secrets $X$ and $Y$ are needed to match the indices.
	\begin{figure*}[t]
		\centering \includegraphics[clip,width=0.8\textwidth,height=0.23\textwidth]{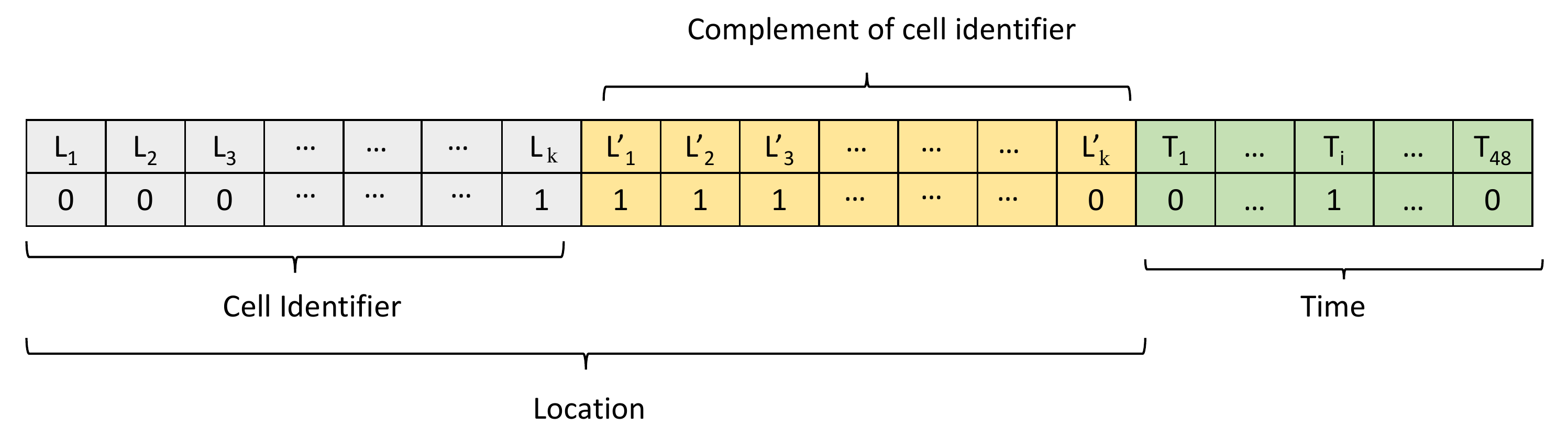}\vspace{-5mm}
		\vspace{2mm}
		\caption{Binary vector of each cell.}
		\label{fig:binary} \vspace{-2mm}
		
	\end{figure*}
	
	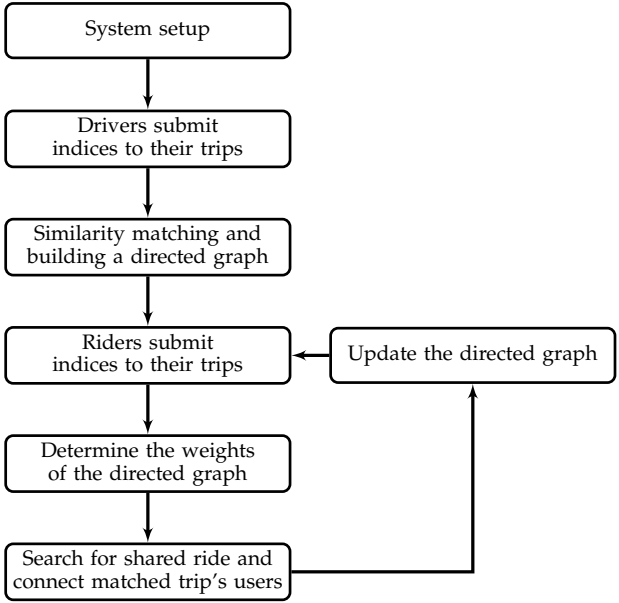
\begin{figure}[!t]
		\centering \scalebox{0.72}{
		
		\tikzstyle{block} = [rectangle, draw, fill=white,     text width=15em, text centered, rounded corners, minimum height=3em, line width=0.5mm] \tikzstyle{line} = [draw, line width=0.6mm, -latex']
		\tikzset{   font={\fontsize{11pt}{12}\selectfont}}
		
		\begin{tikzpicture}[node distance = 2cm, auto]
			\node [block] (init) {System setup};
			\node [block, below of=init] (collect1) {Drivers submit indices to their trips};
			\node [block, below of=collect1] (similarity) {Similarity matching and building a directed graph};
			\node [block, below of=similarity] (collect2) {Riders submit indices to their trips};
			\node [block, right of=collect2, node distance=6cm] (update) {Update the directed graph};
			\node [block, below of=collect2] (decide) {Determine the weights of the directed graph};
			\node [block, below of=decide] (perform) {Search for shared ride and connect matched trip's users};
			\path [line] (init) -- (collect1);
			\path [line] (collect1) -- (similarity);
			\path [line] (similarity) -- (collect2);
			\path [line] (collect2) -- (decide);
			\path [line] (decide) -- (perform);
			\path [line] (update) -- (collect2);
			\path [line] (perform) -| (update);
		\end{tikzpicture}
		
		} \caption{Flowchart of the TRS ridesharing organization scheme.}
		\label{fig:fig132222}
		\end{figure}
	After that, the TOS starts to organize the shared rides as follows.
	The TOS first measures the similarity between the indices of the riders' and the drivers' pick-up time by computing a \emph{dot product} operation between $I_R^{(t)}$ and $I_D^{(t)}$.
	Since only one bit is set to one in the time vector, the dot product result is one if and only if the driver and the rider start their trips within the same time slot.
	In this case, the TOS measures the similarity between the rider and the driver pick-up indices ($I_R^{(C_p)}$ and $I_{D}^{(C_p)}$) as follows.
	\begin{align} \label{eq:IR.ID} \hspace*{-3mm}
		\lefteqn{I_{R}^{(C_p)} X^{\text{-1}} \cdot Y I_{D}^{(C_p)} = }  \nonumber \\
		&\ \ \ \ \ \ \ \ q'^{(C_p)}E_{N}A_{N}p'^{(C_p)}+q'^{(C_p)}E_{N}B_{N}p'^{(C_p)} \nonumber \\
		&\ \ \ \ \ \ +q'^{(C_p)}F_{N}A_{N}p'^{(C_p)}+q'^{(C_p)}F_{N}B_{N}p'^{(C_p)}  \nonumber\\
		&\ \ \ \ \ \ +q''^{(C_p)}G_{N}C_{N}p''^{(C_p)}+q''^{(C_p)}G_{N}D_{N}p''^{(C_p)}  \nonumber\\
		&\ \ \ \ \ \ +q''^{(C_p)}H_{N}C_{N}p''^{(C_p)}+q''^{(C_p)}H_{N}D_{N}p''^{(C_p)}\\
		&\ \ \ \ \ \ =q'^{(C_p)}(E_{N}+F_{N})(A_{N}+B_{N})p'^{(C_p)}  \nonumber\\
		&\ \ \ \ \ \ +q''^{(C_p)}(G_{N}+H_{N})(C_{N}+D_{N})p''^{(C_p)}  \nonumber\\
		&\ \ \ \ \ \ =q'^{(C_p)}p'^{(C_p)}+q''^{(C_p)}p''^{(C_p)}  \nonumber\\
		&\ \ \ \ \ \ =Q^{(C_p)} \cdot P^{(C_p)}  \nonumber
	\end{align}
	Since the rider's pick-up location index $I_R^{(C_p)}$ stores only one cell, the Bloom filter contains exactly $\alpha$ bits set to one. These $\alpha$ bits correspond to the $\alpha$ locations pointed by the hash functions used by the Bloom filter.
	Therefore, if the measured similarity  between $I_R^{(C_p)}$ and $I_D^{(C_p)}$ is exactly $\alpha$, then, the rider's pick-up location lies within the driver's pick-up area. On the other hand, if the measured similarity is less than $\alpha$, then the rider's pick-up location lies outside the driver's pick-up area, and thus ridesharing is not possible between this driver and that rider and TOS has to try other users. On the other hand, if the pick-up time and location are matched, the TOS proceeds based on the requested ridesharing case(s), as follows.
	
	

	\begin{itemize}	
		\item \textit{Organizing MP/MD shared ride}.
		As shown in \autoref{fig:cases}(a), the TOS  measures the similarity of the \textit{rider's drop-off} index $I_R^{(C_d)}$ and the \textit{driver's drop-off} index $I_D^{(C_d)}$. If the result is $\alpha$, then an MP/MD ridesharing is possible.
		
		\item \textit{Organizing MP/RD shared ride}.
		As shown in \autoref{fig:cases}(b), the TOS measures the similarity of the \textit{rider's drop-off} index $I_R^{(C_d)}$ and the \textit{driver's route} index $I_D^{(r)}$. If the result is $\alpha$, then an MP/RD ridesharing is possible.

		\item \textit{Organizing MP/ED shared ride}.
		As shown in \autoref{fig:cases}(c), the TOS measures the similarity of the \textit{driver's drop-off} index $I_D^{(C_d)}$ and the \textit{rider's route} index $I_R^{(r)}$. If the result is $\alpha$, then an MP/ED ridesharing is possible.

	\end{itemize}
	
	Consider that the driver requests to share rides with $N_{r}$ riders.After TOS finds at most $N_{r}$ riders who can share ride with the driver, it connects them.

	
	\subsection{Transferable Ride Sharing Organization} \label{Transferable Ride Sharing Organization Scheme}
	
	\subsubsection{Overview \label{Overview2} }
	The proposed privacy-preserving TRS scheme uses our modified version of the kNN encryption scheme \cite{nabil} and  a modified version of Dijkstra shortest path algorithm.
	\autoref{fig:fig132222} shows flowchart for the TRS scheme. First, each driver creates an encrypted ridesharing offer that includes the individual cells on his trip and sends the offer to the TOS. The TOS should first use the drivers' offers to build a directed graph required for organizing shared rides. Each rider creates an encrypted ridesharing request that includes the pick-up and drop-off cell IDs.  In addition, the TOS determines the weights of the directed graph by using the ridesharing preferences sent by the rider. The TOS uses the riders' requests to search the directed graph to find the path that can satisfy the riders' requirements without learning any location information. Finally, TOS connects the
	drivers and riders. In addition, TOS should make any necessary updates to the directed graph, e.g., if an offer is served, its cells should be removed from the graph. The graph update process does not need much computations and does not require building the whole graph again.

	\subsubsection{Data Representation\label{Data-Representation.-As}}

	As in NRS scheme, in TRS, the ridesharing area is divided
	into geographic regions, called cells, each having a unique identifier $C_i$,
	as illustrated in \autoref{fig:mapa}.
	Unlike NRS, each cell in the driver's route is represented \emph{individually}  by a binary vector that has the location and time components as shown in \autoref{fig:binary}. The location is represented by the cell identifier while the time component is the expected time the user will be in the cell. We define $k$ as the number of bits that are needed to represent each cell identifier in the ridesharing area. For the location component in the cell vector, $2k$ bits are used to represent the location cell identifier in binary and its complement as shown if \autoref{fig:binary}. The complement value is used so that the number of common ones between any two spatially matched cell vectors is fixed value, $k$. For the time component in the cell vector, $\text{\ensuremath{\ell}}$ bits are used to represent the whole day (i.e., 24 hours), where each bit represents a specific time interval. Each user should set the bit corresponding to the time interval in which he/she will be in the cell. Hence, any two spatially and temporally matched cell vectors should have $k+1$ common ones.

	\subsubsection{Driver's Offer Submission}
	\vspace{-2mm}
	
	\label{Data Submitted by the Primary User} For each cell $C_i$ in a
	driver's route, the driver creates a binary vector of length $n=2k+\text{\ensuremath{\ell}}$
	and encrypts it using the modified kNN encryption scheme to get an index vector for each cell in his
	trip's route. For each cell, the driver computes two indices ($I_{D^+}^{(C_i)}$ and $I_{D^-}^{(C_i)}$).  The first index $I_{D^+}^{(C_i)}$ is encrypted using $\mathcal{DTSK}$ and is used to enable riders' requests to be matched with the driver cell data while the second index $I_{D^-}^{(C_i)}$ is encrypted using $\mathcal{RTSK}$ and is used to enable the TOS to match all the drivers' trip data to generate a directed graph required for searching operations.
	For generating $I_{D^+}^{(C_i)}$, the driver uses the binary vector $S_{T}$ to split each binary cell's vector denoted as $P^{(C_i)}$ into two
	vectors $p'^{(C_i)}$ and $p''^{(C_i)}$ of the same size. If the $j^{th}$ bit of $S_{T}$ is zero, then,  $p'^{(C_i)}(j)$ and $p''^{(C_i)}(j)$ are set similar to $P^{(C_i)}(j)$, while if it is one, $p'^{(C_i)}(j)$ and $p''^{(C_i)}(j)$ are set to two random numbers such that their summation is equal to $P^{(C_i)}(j)$.
	After splitting $P^{(C_i)}$, the driver uses $p'^{(C_i)},p''^{(C_i)}$ and
	the key set $\mathcal{DTSK}$ to compute the index $I_{D^+}^{(C_i)}$ as follows.
	\begin{equation}\label{eq:ID_T1}\hspace*{-3mm}
	\begin {split}
	I_{D^+}^{(C_i)}=\Big[
	&Z^{\text{-1}}T_{1}^{\text{-1}}A_{T}p'^{(C_i)};\ Z^{\text{-1}}T_{2}^{\text{-1}}B_{T}p'^{(C_i)};\ Z^{\text{-1}}T_{3}^{\text{-1}}A_{T}p'^{(C_i)};\\
	&Z^{\text{-1}}T_{4}^{\text{-1}}B_{T}p'^{(C_i)};\ Z^{\text{-1}}T_{5}^{\text{-1}}C_{T}p''^{(C_i)};\ Z^{\text{-1}}T_{6}^{\text{-1}}D_{T}p''^{(C_i)};\\
	&Z^{\text{-1}}T_{7}^{\text{-1}}C_{T}p''^{(C_i)};\ Z^{\text{-1}}T_{8}^{\text{-1}}D_{T}p''^{(C_i)}\Big]
	\end{split}
	\end{equation}
	
	where  $I_{D^+}^{(C_i)}$ is a column vector of size $8$ elements, and each element is a column vector of size $n$. For generating $I_{D^-}^{(C_i)}$, the binary vector $S_{T}$ is used to split the binary cell's vector denoted as $Q^{(C_i)}$ into two  vectors $q'^{(C_i)}$ and $q''^{(C_i)}$ of the same size.
	The splitting process of $Q^{(C_i)}$ is opposite to that of $P^{(C_i)}$.
	If the $j^{th}$ bit of $S_{T}$ is one, then,  $q'^{(C_i)}(j)$ and $q''^{(C_i)}(j)$ are set similar to $Q^{(C_i)}(j)$, while if it is zero, $q'^{(C_i)}(j)$ and $q''^{(C_i)}(j)$ are set to two random numbers
	such that their summation is equal to $Q^{(C_i)}(j)$.
	Then, the driver can compute the encrypted index $I_{D^-}^{(C_i)}$ using the key set $\mathcal{RTSK}$ and the vectors $q'^{(C_i)}$ and $q''^{(C_i)}$ as follows
	\begin{equation}\label{eq:ID_T2}\hspace*{-3mm}
		\begin {split}
	I_{D^-}^{(C_i)}=\Big[
	&q'^{(C_i)}E_{T}T_{1}W,\ q'^{(C_i)}E_{T}T_{2}W,\ q'^{(C_i)}F_{T}T_{3}W\\
	&q'^{(C_i)}F_{T}T_{4}W,\ q''^{(C_i)}G_{T}T_{5}W,\ q''^{(C_i)}G_{T}T_{6}W,\\
	&q''^{(C_i)}H_{T}T_{7}W,\ q''^{(C_i)}H_{T}T_{8}W\Big]
	\end {split}
	\end{equation}
	where $I_{D^-}^{C_i}$ is a row vector of size $8$ elements, and each element is a row vector of size $n$. Each driver should encrypt and send the indices of the individual cells of his route to the TOS. Other information such as the number of riders who can share the trip with, and his encrypted contact information should be sent to the TOS along with anonymous signature.

	\subsubsection{Building A Directed Graph\label{subsec:Similarity-Matching-and}}

	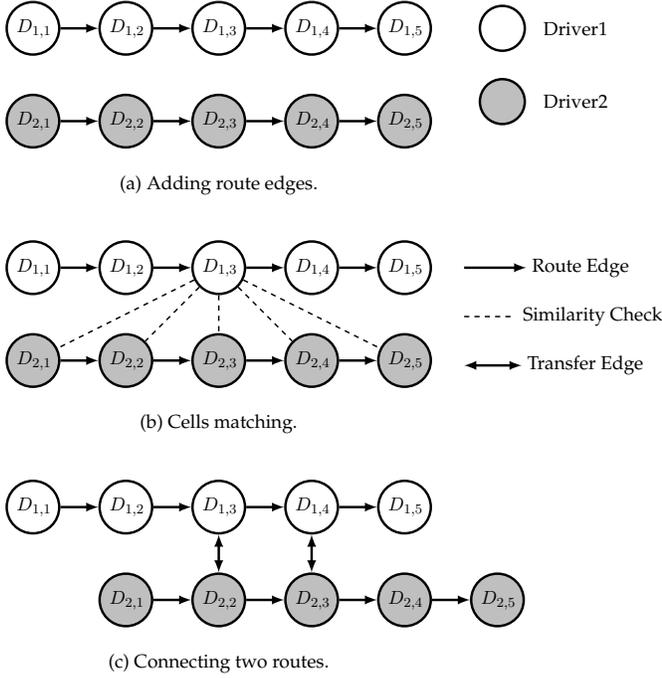
\begin{figure}[!t]
		\scalebox{0.65}{
			
			\tikzstyle{circle1} = [circle, draw, fill=white,     text width=2em, text centered, rounded corners, minimum height=1.5em, line width=0.5mm]
			\tikzstyle{circle2} = [circle, draw, fill=lightgray,     text width=2em, text centered, rounded corners, minimum height=1.5em, line width=0.5mm]
			\tikzstyle{line} = [draw, line width=0.5mm, -latex]
			\tikzstyle{dotline} = [dashed, line width=0.3mm]
			\tikzstyle{doubleline} = [latex-latex,line width=0.5mm]
			\tikzset{   font={\fontsize{11pt}{12}\selectfont}}
			
			\begin{tikzpicture}[node distance = 2cm, auto]
			\node [circle1] (p11) {$D_{1,1}$};
			\node [circle1,right of=p11,node distance=1.9cm] (p12) {$D_{1,2}$};
			\node [circle1,right of=p12,node distance=1.9cm] (p13) {$D_{1,3}$};
			\node [circle1,right of=p13,node distance=1.9cm] (p14) {$D_{1,4}$};
			\node [circle1,right of=p14,node distance=1.9cm] (p15) {$D_{1,5}$};

			\node [circle2, below of= p11,node distance=1.9cm] (p21) {$D_{2,1}$};
			\node [circle2,right of=p21,node distance=1.9cm] (p22) {$D_{2,2}$};
			\node [circle2,right of=p22,node distance=1.9cm] (p23) {$D_{2,3}$};
			\node [circle2,right of=p23,node distance=1.9cm] (p24) {$D_{2,4}$};
			\node [circle2,right of=p24,node distance=1.9cm] (p25) {$D_{2,5}$};
			\node [below of=p23,node distance=1.3cm] (p26) {(a) Adding route edges.};
			
			\node [circle1,right of=p15,node distance=2cm] (xx){};
			\node [right of=xx,node distance=1.5cm] {Driver1};		
			\node [circle2,below of=xx,node distance=1.5cm] (yy){};
			\node [right of=yy,node distance=1.5cm] {Driver2};
			
			\path [line] (p11) -- (p12);
			\path [line] (p12) -- (p13);
			\path [line] (p13) -- (p14);		
			\path [line] (p14) -- (p15);
			
			\path [line] (p21) -- (p22);
			\path [line] (p22) -- (p23);
			\path [line] (p23) -- (p24);		
			\path [line] (p24) -- (p25);

			\node [circle1,below of= p21,node distance=3cm] (p31) {$D_{1,1}$};
			\node [circle1,right of=p31,node distance=1.9cm] (p32) {$D_{1,2}$};
			\node [circle1,right of=p32,node distance=1.9cm] (p33) {$D_{1,3}$};
			\node [circle1,right of=p33,node distance=1.9cm] (p34) {$D_{1,4}$};
			\node [circle1,right of=p34,node distance=1.9cm] (p35) {$D_{1,5}$};

			\node [circle2, below of= p31,node distance=1.9cm] (p41) {$D_{2,1}$};
			\node [circle2,right of=p41,node distance=1.9cm] (p42) {$D_{2,2}$};
			\node [circle2,right of=p42,node distance=1.9cm] (p43) {$D_{2,3}$};
			\node [circle2,right of=p43,node distance=1.9cm] (p44) {$D_{2,4}$};
			\node [circle2,right of=p44,node distance=1.9cm] (p45) {$D_{2,5}$};
			\node [below of=p43,node distance=1.3cm] (p46) {(b) Cells matching.};

			\node [right of=p35,node distance=1.1cm] (aa){};
			\node [right of=aa,node distance=2.5cm] (bb){Route Edge};
			\path [line] (aa) -- (bb);

			\node [below of=aa,node distance=1cm] (aa2){};
			\node [right of=aa2,node distance=2.7cm] (bb2){\ Similarity Check};
			\draw [dotline] (aa2) -- (bb2);	
			
			\node [below of=aa2,node distance=1cm] (aa1){};
			\node [right of=aa1,node distance=2.6cm] (bb1){Transfer Edge};
			\draw [doubleline] (aa1) -- (bb1);
			

			\path [line] (p31) -- (p32);
			\path [line] (p32) -- (p33);
			\path [line] (p33) -- (p34);		
			\path [line] (p34) -- (p35);
			
			\path [line] (p41) -- (p42);
			\path [line] (p42) -- (p43);
			\path [line] (p43) -- (p44);		
			\path [line] (p44) -- (p45);
			
			\draw [dotline] (p33) -- (p41);
			\draw [dotline] (p33) -- (p42);
			\draw [dotline] (p33) -- (p43);
			\draw [dotline] (p33) -- (p44);
			\draw [dotline] (p33) -- (p45);

			\node [circle1, below of= p41,node distance=3cm] (p51) {$D_{1,1}$};
			\node [circle1,right of=p51,node distance=1.9cm] (p52) {$D_{1,2}$};
			\node [circle1,right of=p52,node distance=1.9cm] (p53) {$D_{1,3}$};
			\node [circle1,right of=p53,node distance=1.9cm] (p54) {$D_{1,4}$};
			\node [circle1,right of=p54,node distance=1.9cm] (p55) {$D_{1,5}$};
			
			\node [circle2, below of= p52,node distance=1.9cm] (p61) {$D_{2,1}$};
			\node [circle2,right of=p61,node distance=1.9cm] (p62) {$D_{2,2}$};
			\node [circle2,right of=p62,node distance=1.9cm] (p63) {$D_{2,3}$};
			\node [circle2,right of=p63,node distance=1.9cm] (p64) {$D_{2,4}$};
			\node [circle2,right of=p64,node distance=1.9cm] (p65) {$D_{2,5}$};
			\node [below of=p62,node distance=1.3cm] (p66) {(c) Connecting two routes.};

			\path [line] (p51) -- (p52);
			\path [line] (p52) -- (p53);
			\path [line] (p53) -- (p54);		
			\path [line] (p54) -- (p55);

			\path [line] (p61) -- (p62);
			\path [line] (p62) -- (p63);
			\path [line] (p63) -- (p64);		
			\path [line] (p64) -- (p65);
			
			\draw [doubleline] (p53) -- (p62);
			\draw [doubleline] (p54) -- (p63);
			
			\end{tikzpicture}
			
		} \caption{Similarity measurement and building the graph. (a) Two drivers routes with route edges added. (b) The cell matching between $D_{1,3}$ cell and all cells of driver two's route. (c) The connected graph of the two drivers after adding bidirectional edges between matched cells. Note that, $D_{1,3}$ and $D_{2,2}$ have the same physical location, similarly, $D_{1,4}$ and $D_{2,3}$.}
		\label{fig:buildingraph}
	\end{figure}
	After TOS receives the cell indices of drivers'
	routes, it constructs a directed graph that contains all the cells of all the drivers without knowing the exact spatial locations of the cells to preserve privacy.
	
	Given driver $h$ route, where cell $j$ on his/her route is denoted as $D_{k,j}$, the TOS builds the directed graph by measuring the similarity of cells' indices. To do this, the following two types of graph edges are used:
	\begin{enumerate}
	\item \textbf{Route Edges.} These edges are unidirectional and connects
	different cells that belong to the same driver.
	\item \textbf{Transfer Edges.} These edges are bidirectional and connects
	the matched cells of different drivers. In these cells, the drivers will be in the same time at the same location and thus a rider can transfer from one driver to another.
	\end{enumerate}
	
	The TOS adds a route edge between any two consecutive cells
	$D_{h,j}$ and $D_{h,j+1}$ of driver $h$. As shown
	in \autoref{fig:buildingraph}(a), there are two drivers each
	having five cells in his route and the route edges are added between the same driver cells. After adding route edges, TOS measures the similarity of every driver's cell to all other drivers' cells in order to add transfer edges to the graph. \autoref{fig:buildingraph}(b)
	shows that TOS measures the similarity of driver one's cell three  ($D_{1,3}$)  and  all the indices of driver two.  If any of the matching result is exactly equal to $k+1$, it means that both drivers will be in the same cell at the same time. In that case, TOS adds a bidirectional edge (transfer edge)
	between the two cells. As shown in \autoref{fig:buildingraph}(c), TOS finishes adding all the
	transfer edges to connect the two routes. After connecting the two routes, a third driver's route can be added to \autoref{fig:buildingraph}(c) graph using the same procedure and the directed graph is completed after connecting the routes of all drivers. Note that, even if a graph is constructed and a new driver sends a route, it is easy to add this route to the constructed graph without building the whole graph.

	\subsubsection{Rider's Request Submission}
	For the rider, his trip data contains only two cells; one cell for pick-up $C_p$ and another cell for drop-off $C_d$. Using the key set $\mathcal{RTSK}$, the rider can compute two indices $I_{R}^{(C_p)}$ and $I_{R}^{(C_d)}$ and send them to the TOS.
	Also, the rider should encrypt his contact information and sends it to the TOS along with an anonymous signature.
	\subsubsection{Organization of Shared Rides}

	\textbf{Graph Weights Based on Rider's Preferences.}
	Since riders have different requirements and needs, our scheme
	allows them to prescribe their preferences in the shared rides.  The riders' preferences are used by the TOS to determine the graph's weights so that the optimal search result is returned to the rider. The number of cells traveled by the rider and the number of transfers done by the rider are used to prescribe the rider's preferences. Riders' preferences are given as follows.
	\begin{enumerate}
	\item \textit{Minimum number of cells ($Min_{c}$).} In this preference, rider requires a trip with the minimum number of cells (shortest distance). If TOS finds different
	trips that can satisfy $Min_{c}$, it returns
	the trip that has the least number of transfers.
	\item \textit{Maximum number of cells ($Max_{c}$).} In this preference,
	rider requires TOS to return a trip that
	has a number of cells that are less than a threshold value $Max_{c}$. If TOS finds different
	trips that can satisfy $Max_{c}$, it returns the trip that has the least number of transfers.
	\item \textit{Minimum number of transfers ($Min_{t}$).} In this preference,
	rider requires a trip that has the minimum number of transfers. This
	option may be preferable by elder people or the people
	with disability on wheelchairs. If TOS finds different
	trips that can satisfy $Min_{t}$, it returns
	the trip that has the least number of cells.
	\item \textit{Maximum number of transfers ($Max_{t}$).} In this preference,
	the rider requires TOS to return only the trips that
	has a number of transfers that is less than a threshold value $Max_{t}$. If TOS finds different
	trips that can satisfy $Max_{t}$, it returns the trip that has the least number of cells.
	\end{enumerate}
	
	In addition, any combinations of the aforementioned preferences can
	be used such as, minimum number of cells and transfers ($Min_{c,t}$),
	minimum number of transfers and maximum number of cells ($Min_{t},Max_{c}$),
	Minimum number of cells and maximum number of transfers ($Min_{c},Max_{t}$),
	and maximum number of cells and maximum number of
	transfers ($Max_{c,t}$).

	When TOS receives the rider's data, it uses the rider's preference to set the weight of each edge in the directed graph. For $Min_{c}$ and $Max_{c}$, the weights of route edges are set to one and the weights for transfer edges should be zero as they occur within the same cell, however, in order to eliminate zero weight cycles in the graph, a small value, $\epsilon$, is used. On the other hand, for $Min_{t}$ and $Max_{t}$, the weight of transfer edges is one, while the weight of route edges is $\epsilon$.  For the other combinations of the aforementioned preferences, a weighting process similar to that described for the above two preferences is used. After determining the weight
	of each edge in the graph based on the rider's preference, the TOS is ready to run the search algorithm.
	
\begin{figure}[!t]
		\scalebox{0.63}{
			
			\tikzstyle{circle1} = [circle, draw, fill=white,     text width=2em, text centered, rounded corners, minimum height=1.5em, line width=0.5mm]
			\tikzstyle{circle2} = [circle, draw, fill=lightgray,     text width=2em, text centered, rounded corners, minimum height=1.5em, line width=0.5mm]
			\tikzstyle{circle3} = [circle, draw, fill=mintgreen,     text width=2em, text centered, rounded corners, minimum height=1.5em, line width=0.5mm]		
			\tikzstyle{line} = [draw, line width=0.5mm, -latex]
			\tikzstyle{dotline} = [dashed, line width=0.3mm]
			\tikzstyle{doubleline} = [latex-latex,line width=0.5mm]
			\tikzset{   font={\fontsize{11pt}{12}\selectfont}}
			
			\begin{tikzpicture}[node distance = 2cm, auto]
			\node [circle1,double] (p11) {$D_{1,1}$};
			\node [below of= p11,node distance=1cm] (source) {Source};
			\node [circle1,right of=p11,node distance=1.8cm] (p12) {$D_{1,2}$};
			\node [circle1,right of=p12,node distance=1.8cm] (p13) {$D_{1,3}$};
			\node [circle1,right of=p13,node distance=1.8cm] (p14) {$D_{1,4}$};

			\node [circle2, below of= p13,node distance=2.2cm] (p21) {$D_{2,1}$};
			\node [circle2,right of=p21,node distance=1.8cm] (p22) {$D_{2,2}$};
			\node [circle2,right of=p22,node distance=2.75cm] (p23) {$D_{2,3}$};
			
			\node [circle2,double,right of=p23,node distance=2.75cm] (p25) {$D_{2,4}$};

			\node [circle3,below of= p21,node distance=2.2cm] (p31) {$D_{3,1}$};
			\node [circle3,right of=p31,node distance=1.8cm] (p32) {$D_{3,2}$};
			\node [circle3,right of=p32,node distance=1.8cm] (p33) {$D_{3,3}$};
			\node [circle3,right of=p33,node distance=1.8cm] (p34) {$D_{3,4}$};
			\node [circle3,double,right of=p34,node distance=1.8cm] (p35) {$D_{3,5}$};

			\node [below of= source,node distance=4.8cm,anchor=west] (t1) {\Large
				$Min_c : {D_{1,1},D_{1,2},D_{1,3},D_{2,1},D_{2,2},D_{2,3},D_{2,4}}$};
			\node [below of= source,node distance=5.4cm,anchor=west] (t2) {\Large
				$Min_t : {D_{1,1},D_{1,2},D_{1,3},D_{2,1},D_{2,2},D_{2,3},D_{2,4}}$};
			\node [below of= source,node distance=6cm,anchor=west] (t3) {\Large
				$Min_t : {D_{1,1},D_{1,2},D_{1,3},D_{3,1},D_{3,2},D_{3,3},D_{3,4},D_{3,5}}$};
			\node [below of= source,node distance=6.6cm,anchor=west] (t4) {\Large
				$Min_{c,t} : {D_{1,1},D_{1,2},D_{1,3},D_{2,1},D_{2,2},D_{2,3},D_{2,4}}$};
			
			\path [line] (p11) -- (p12);
			\path [line] (p12) -- (p13);
			\path [line] (p13) -- (p14);

			\path [line] (p21) -- (p22);
			\path [line] (p22) -- (p23);
			\path [line] (p23) -- (p25);		
			
			\path [line] (p31) -- (p32);
			\path [line] (p32) -- (p33);
			\path [line] (p33) -- (p34);		
			\path [line] (p34) -- (p35);
			
			\path [line] (p34) -- (p35);
			
			\draw [doubleline] (p13) -- (p21);	
			\draw [doubleline] (p21) -- (p31);	
			\draw [doubleline] (p22) -- (p32);
			\draw [doubleline] (p25) -- (p35) node[midway]{\ Destinations};
			\draw [doubleline] (p13) .. controls +(-1.5,-1) and +(-1.5,1).. (p31);

			\node [right of=p14,node distance=2.5cm] (aa){};
			\node [right of=aa,node distance=2.5cm] (bb){Route Edge};
			\path [line] (aa) -- (bb);

			\node [below of=aa,node distance=0.8cm] (aa1){};
			\node [right of=aa1,node distance=2.6cm] (bb1){Transfer Edge};
			\draw [doubleline] (aa1) -- (bb1);
			\end{tikzpicture}
			
		} \caption{Examples of ridesharing preferences assuming three drivers and one rider request from node $D_{1,1}$ with either node $D_{2,4}$ or $D_{3,5}$ as destination. Note that, two solution exist for $Min_t$ with different number of cells.}
		\label{fig:ridecases}
	\end{figure}
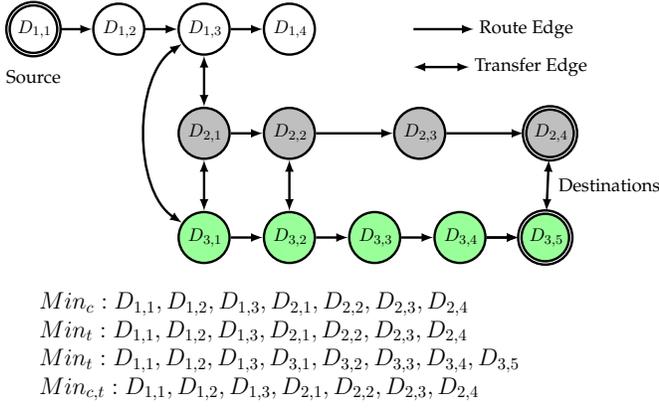

	\autoref{fig:ridecases} gives an example for the different options supported by
	the proposed scheme that satisfy all the requirements and needs of the riders. In this figure, three drivers' routes are connected through a directed graph. It assumes that one
	rider requests a ride from the source node $D_{1,1}$  to the destination node  $D_{2,4}$ or $D_{3,5}$. Note that, nodes  $D_{2,4}$ and $D_{3,5}$ have the same physical cell. If the rider requests $Min_c$, the TOS returns $\{D_{1,1},D_{1,2},D_{1,3},D_{2,1},D_{2,2},D_{2,3},D_{2,4}\}$, for $Min_t$ the result is either $\{D_{1,1},D_{1,2},D_{1,3},D_{2,1},D_{2,2},D_{2,3},D_{2,4}\}$ or $\{D_{1,1},D_{1,2},D_{1,3},D_{3,1},D_{3,2},D_{3,3},D_{3,4},D_{3,5}\}$ because both of these paths have only one transfer, and for $Min_{c,t}$ the TOS returns $\{D_{1,1},D_{1,2},D_{1,3},D_{2,1},D_{2,2},D_{2,3},D_{2,4}\}$. \\
	
	\textbf{Search the Graph.}
	To organize shared rides, TOS should first search the directed graph to find the route and hence the drivers that the rider will share the ride with. Compared to other graph searching algorithms that target the shortest paths between cells in a weighted graphs, Dijkstra's algorithm is the fastest single-source multi destination shortest path algorithm with searching complexity of $O(v\hspace{1mm}log(v))$ \cite{ri24}, where $v$ is the number of vertices in the graph.
	
	We modified Dijkstra's searching algorithm \cite{ri24} to enable TOS find all shortest paths with equal cost which the
	existing Dijkstra's algorithm cannot find. The original algorithm returns only one shortest path,
	whereas in our scheme, some riding preferences need the whole set of shortest paths resulted from a certain request.
	For example, in $Min_{c,t}$, we need to know the whole set of the paths
	that satisfy the minimum number of cells and the whole set of
	paths that satisfy the minimum number of transfers to take the intersection
	between them to get the paths that can achieve minimum numbers of
	cells and transfers. Given a rider's source and destination nodes, the modified algorithm can find all the shortest paths connecting them on the graph. The modified algorithm is illustrated in Algorithm \ref{Ag1}.
	
	\begin{algorithm}[t]
		\SetAlgoLined \KwData{Graph,source} \KwResult{dist,pred}
		PQ=PiriorityQueue \\
		 \For{each vertex v in Graph} {
			$dist[v]\gets \infty$\\
			 $pred[v]\gets[\ ]$\\
		 } $dist[source]\gets0$ \\
		 $PQ.insert(source,0)$\\
		 \While{not PQ.empty()} { $u\gets PQ.pop()$\\
		 \For{each $neighbor$ $v$ of $u$} { \uIf{$dist[v]>dist[u]+Length(u,v)$}
		{ $dist[v]\gets dist[u]+Length(u,v)$\\
		 $PQ.insert(v,dist(v))$\\
		 $pred[v]\gets u$ } \uElseIf{$dist[v]==dist[u]+Length(u,v)$}
		{ $pred[v].add(u)$ } }
		} \caption{Modified Dijkstra Algorithm \label{Ag1} }
	\end{algorithm}
		
	In order to organize shared rides that can satisfy the preference
	of the riders, TOS matches the startup cell of the
	rider to all the cells in the graph to determine the possible source cells. The same procedure should be performed using the rider's destination
	cell to find the possible destination cells. Note that, possible source/destination cells have the same physical location, however, they belong to different drivers.
As an example for similarity matching, consider the TOS measures the similarity of an arbitrary cell $C_a$ encrypted using $\mathcal{RTSK}$ and has an index $I_R^{(C_a)}$  and another cell $C_b$ encrypted using $\mathcal{DTSK}$ and has an index $I_D^{(C_b)}$. The similarity measurement can be computed as follows
	
	\begin{align}  \label{eq:IRID2}
		\lefteqn{I_{R}^{(C_b)} W^{\text{-1}} \cdot Z I_{D}^{(C_b)} =}  \nonumber \\
		&q'^{(C_a)}E_{T}A_{T}p'^{(C_b)}+q'^{(C_a)}E_{T}B_{T}p'^{(C_b)}  \nonumber \\
		&+q'^{(C_a)}F_{T}A_{T}p'^{(C_b)}+q'^{(C_a)}F_{T}B_{T}p'^{(C_b)}  \nonumber \\
		&+q''^{(C_a)}G_{T}C_{T}p''^{(C_b)}+q''^{(C_a)}G_{T}D_{T}p''^{(C_b)}\\
		&+q''^{(C_a)}H_{T}C_{T}p''^{(C_b)}+q''^{(C_a)}H_{T}D_{T}p''^{(C_b)}  \nonumber \\
		&=q'^{(C_a)}(E_{T}+F_{T})(A_{T}+B_{T})p'^{(C_b)}  \nonumber  \\
		&+q''^{(C_a)}(G_{T}+H_{T})(C_{T}+D_{T})p''^{(C_b)}  \nonumber \\
		&=q'^{(C_a)}p'^{(C_b)}+q''^{(C_a)}p''^{(C_b)}  \nonumber \\
		&=Q^{(C_a)} \cdot P^{(C_b)}  \nonumber
	\end{align}

Then, TOS uses the modified Dijkstra's searching algorithm to find all paths between source and destination that satisfy the
	required preferences. Finally, the TOS should return the contact information of the riders/drivers and the ciphertext of the cells of the transfers. In addition, TOS updates the directed graph by removing the edges of the associated offers if they have exhausted their ridesharing capacity.

	\section{Privacy Analysis} \label{sec:privacy_analysis}
	
	In this section, we discuss how our schemes can achieve the privacy-preservation
	goals mentioned in \autoref{design_goals}. For the following formal proof, we followed the same model in \cite{curtmola2011searchable} to
	prove the security of our schemes.
	
		\begin{theorem}
			The TOS can organize shared NRS and TRS rides without learning any sensitive information
			about the trip's data.
		\end{theorem}
		
		\vspace{1mm}\begin{IEEEproof}
			 We provide a proof for the theorem to the NRS scheme and by the same analogy the proof can be used for TRS. First, we define some notations:

			\vspace{2mm} \emph{History.}\textbf{ }For a set of drivers' routes $R$, the history
			is a set of indices $\mathcal{{I}}=\{I^{(r_1)}_{D_{1}},\dots,I^{(r_{n\_d})}_{D_{n\_d}}\}$
			over $R$ and a set of riders pick-up locations $\mathcal{{I_{P}}}=\{I^{(C_{p_1})}_{R_{1}},I^{(C_{p_2})}_{R_{2}},\dots,I^{(C_{p_{k\_r}})}_{R_{k\_r}}\}$,
			denoted as $H=(\mathcal{{I}},\ \mathcal{{I_{P}}})$, where $n\_d$ and $k\_r$ are the number of drivers and riders, respectively.
			
			\vspace{1mm}\emph{Trace.}\textbf{ }A trace reflects the knowledge inferred by
			the TOS over the history $H$, denoted as $Tr(H)$, such as
			the search and access patterns, where $Tr(H)$ is defined over all
			the riders' pick-up locations of $H$, such as $Tr(H)=\{Tr (I^{(C_{p_1})}_{R_{1}}),Tr (I^{(C_{p_2})}_{R_{2}}),\dots,Tr (I^{(C_{p_{k\_r}})}_{R_{k\_r}})\}$
			
			\vspace{1mm}\emph{View.} It represents the perception of the TOS.
			It is the combination of the encrypted history and its trace, denoted as $V(\mathcal{{I}},\mathcal{{I_{P}}},Tr(H))$.
			\vspace{2mm}
			
			Consider a simulator $\mathcal{S}$ that can generate a false view
			$V'$ that is indistinguishable from $V$ by doing the following steps.
			\begin{enumerate}
				\item $\mathcal{S}$ generates a master secret key $sk'=  K'_{NT}$.
				\item $\mathcal{S}$ generates a set of random routes $R'=\{r'_{1},\dots,r'_{n\_d}\}$ such
				that $|r_{i}|=|r'_{i}|$, $1\leq i\leq n\_d$, $r_{i}'=\{C_{1}',C_{2}',\dots\}$, and $|r_{i}|$ is the length of the route.
				\item $\mathcal{S}$ generates a set of pick-up cells $\mathcal{{I_{P}}}'=\{I'^{(C_{p_1})}_{R_{1}},I'^{(C_{p_2})}_{R_{2}},\dots,I'^{(C_{p_{k\_r}})}_{R_{k\_r}}\}$,
				where $I'^{(C_{p_j})}_{R_{j}} \in\{0,1\}^{m}$ is generated as follows by replacing each $C_{pj}$ it with $C_{pj}'$ and generate  $I'^{(C_{p_j})}_{R_{j}}$. Note that $I'^{(C_{p_j})}_{R_{j}}$
				is  a random copy of $I^{(C_{p_j})}_{R_{j}}$.
				\item $\mathcal{S}$ generates a set of $m$-bit vectors denoted as indices
				$\mathcal{{I}}'=\{I'^{(r_1)}_{D_{1}},\dots,I'^{(r_{n\_d})}_{D_{n\_d}}\}$ as follows. For each cell $C_{j}\in city$, if $C_{j}\subset r_{i}$ and $1\leq i\leq n\_d$,
				add $C_{j}'$ to $I'^{(r_i)}_{D_{i}}$. Note that $I'^{(r_i)}_{D_{i}}$ is a
				random copy of $I^{(r_i)}_{D_{i}}$ .
				\item $\mathcal{S}$ generates encrypted indices $\mathcal{{I}}'$
				and trapdoor $\mathcal{{I_{P}}}'$ using secret $sk'$.
			\end{enumerate}
			From the previous construction $\mathcal{S}$ has a history $H'=(\mathcal{{I}}',\ \mathcal{{I_{P}}}')$ with trace $Tr(H')$ similar to $Tr(H)$ such that in no probabilistic polynomial-time (P.P.T.), adversary can distinguish between the two
			views $V(\mathcal{{I}},\mathcal{{I_{P}}},Tr(H))$
			and $V'(\mathcal{{I}'},\mathcal{{I_{P}}}',Tr(H'))$
			with non-negligible advantage where the correctness of the construction
			implies this conclusion. In particular, the indistinguishability follows
			directly from the semantic security of kNN encryption scheme.
		
	\end{IEEEproof}
	In addition, the random numbers used in kNN encryption scheme can make the ciphertext of the same data look different when encrypted
	multiple times. If a malicious user intercepts the ciphertext of another
	user's trip, he cannot decrypt it because the secret key set is needed for
	decryption and each user has a unique key.

	Beside the aforementioned features, the proposed  schemes acheive the following privacy features.
	
	\begin{enumerate}
	\item \textit{Users' anonimity.} By using
	anonymous signature, TOS can learn that a received ridesharing request/offer is sent by a legitimate user without learning his/her identity. Also, by hiding the users' locations, the TOS cannot identify the users from the locations they visit.
	
	\item \textit{Offers/requests matching.} Users and eavesdroppers cannot match drivers' offers to riders' requests. Only the TOS can match the ridesharing offers and requests because the TOS needs to use its secrets as indicated in Equation \eqref{eq:IR.ID} and \eqref{eq:IRID2}.
	
	
	\item \textit{Forward and backward privacy}: By frequently changing the cells' identifiers, the TOS cannot collect side information by matching the indices sent by the users at a given time to the past or future indices.
	
	\item \textit{Requests-users un-linkability.} Given different ridesharing
	requests sent from one user at different times, TOS cannot
	learn if these requests are sent from the same user or not.
	
	\item \textit{Driver-rider pair un-linkability.} If one rider
	shares a ride with one driver, our schemes do not leak any information that can help the TOS to identify the driver-rider pair when they share rides in the future. 
This is because of the use of anonymous signature and because the trip's encrypted data cannot be matched to old data due to changing the cells' identifiers.
	\item \textit{Same request un-linkability.} The encrypted requests of the same trip look different when encrypted at different times even if they have the same trip information. This is due to the random numbers used in the kNN encryption scheme to encrypt the requests.
	
	\end{enumerate}

	\section{Performance Evaluations} \label{sec:performance}
	
	\subsection{Experiment Setup}
	The proposed schemes were implemented using MATLAB, and a server with
	an Intel\textregistered{} Xeon\textregistered{} Processor E5-2420
	@2.2GHZ (2 processors) and 32 GB RAM. In our experiments, we used
	real maps to create a set of trips extracted from the map. First, we used OpenStreetMap project \cite{ri28}
	to get a real map for the city of Nashville, TN, USA as shown
	in \autoref{fig:mapa}. The ridesharing area is 75.5 km $\times$
	33 km. This area is divided into 15,687 cells with a cell size of
	400 m $\times$ 400 m. We used SUMO program \cite{ri29}
	to create real and random routes for 450 users. All the reported results
	are averaged over 30 different runs. In NRS, the Bloom filter size
	is computed to be 2,048 bits such that the false positive probability
	is at most 0.01 assuming at most 60 cells are in trip routes. In
	TRS, the time component $\text{\ensuremath{\ell}}$ is represented by different numbers of bits that are between 25 and 200, while 48 bits are used for NRS.

	\subsection{Non-transferable ridesharing organization}
	
	\label{sec:performance1}
	
	\subsubsection{Performance Metrics}
	
	Three performance metrics are used for comparison and assessment between our
	NRS scheme and the proposed scheme in the literature, complete city representation-based scheme(CCRS) \cite{ri22}.
	\begin{enumerate}
	\item \textit{Search time:} The time needed by TOS to organize shared
	rides.
	\item \textit{Computation overhead:} The time needed by the diver, rider, and TOS to run NRS scheme. Since the users send their
	requests/offers using their mobile phones that may have low computational
	power, reducing the computation overhead on the users is desirable.
	\item \textit{Communication overhead:} The amount of data transmitted during
	the communication between the users and the TOS.
	\end{enumerate}
	
	\subsubsection{Experiment Results}
	\label{Experiment Results1}

	\textbf{Computation Overhead.} \autoref{tab:2} summarizes the computation
	overhead results of NRS versus CCRS. For the modified kNN encryption scheme, the secret key generation takes 198.4 sec for CCRS and
	2.7 sec per user for our scheme. For encryption, the computation of an encrypted request takes 1.26
	sec for CCRS and 0.15 sec for our scheme. Note that, the data vector in NRS is much shorter than that of CCRS, this because the driver's route in CCRS is represented by a vector with size equal to the total number of cells in the city. However in NRS, only route cells identifiers are added to the Bloom filter to be encrypted. Hence, NRS requires less time for inner product operations in comparison to CCRS.
	
	\begin{figure}[!t]
	\centering \setlength{\abovecaptionskip}{0mm} \setlength{\belowcaptionskip}{-2mm}
	\includegraphics[clip,scale=0.39]{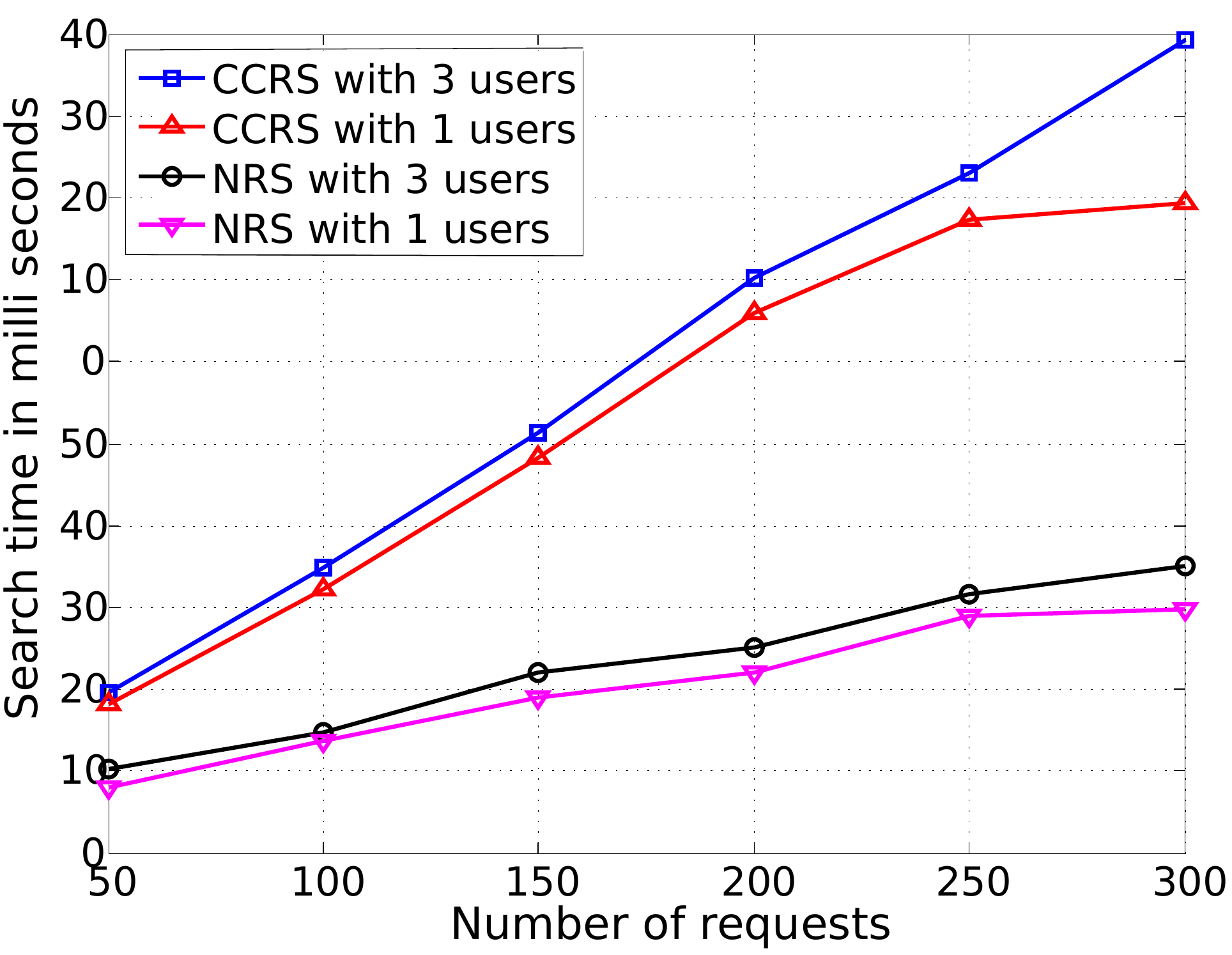}
	\caption{Search time versus number of riders' requests.}
	\vspace{1mm}
	\label{fig:fig11}
	\end{figure}
	\begin{table}[!t]
		\caption{Computation overhead.{\small{}\label{tab:2} }}
		\center
		\begin{tabular}{cccc}
		\toprule
		\multicolumn{2}{c}{} &  CCRS &  NRS\tabularnewline
		\midrule
		\midrule
		Modified kNN  & \multirow{1}{*}{Key generation} & 198.42 sec & 2.7 sec\tabularnewline
		\cmidrule{2-4}
		Encryption Scheme{\small{} } & Index generation& 1.26 sec & 0.15 sec\tabularnewline
		\bottomrule
		\end{tabular}
	\end{table}
	\textbf{Search Time.} \autoref{fig:fig11} gives the search time
	in milliseconds versus the number of requests. Two cases are considered.
	In the first case, the driver intends to share rides with only
	one rider. In the second case, driver intends to share ride
	with up to three riders. We also assumed that the
	driver's preferences are MP/MD, MP/RD, and
	MP/ED in order. 
	
	The figure shows that \textit{the search time in our scheme can be less than half that of CCRS}.  From the figure, when the number of requests is 50, 150 and 300,  NRS achieves a 50\%, 60\% and 65\% reduction, respectively compared to CCRS.
	This reduction can be attributed to the fact that in CCRS all the
	15,687 cells must be represented in the pick-up, drop-off and route
	vectors with 15,687 bits for each vector, while in NRS, only the cells
	of pick-up/drop-off and routes are added to Bloom filters with
	a much smaller size. 
	Consequently, CCRS requires more time for inner product operations to match
	users encryption to organize rides. The figure also shows that
	an increase in the number of rider requests increases the
	search time in the two schemes, but compared to CCRS, NRS requires
	much less search time.
	
	\begin{figure}[t]
		\centering \setlength{\abovecaptionskip}{0mm} \setlength{\belowcaptionskip}{-2mm}
		\includegraphics[clip,scale=0.45]{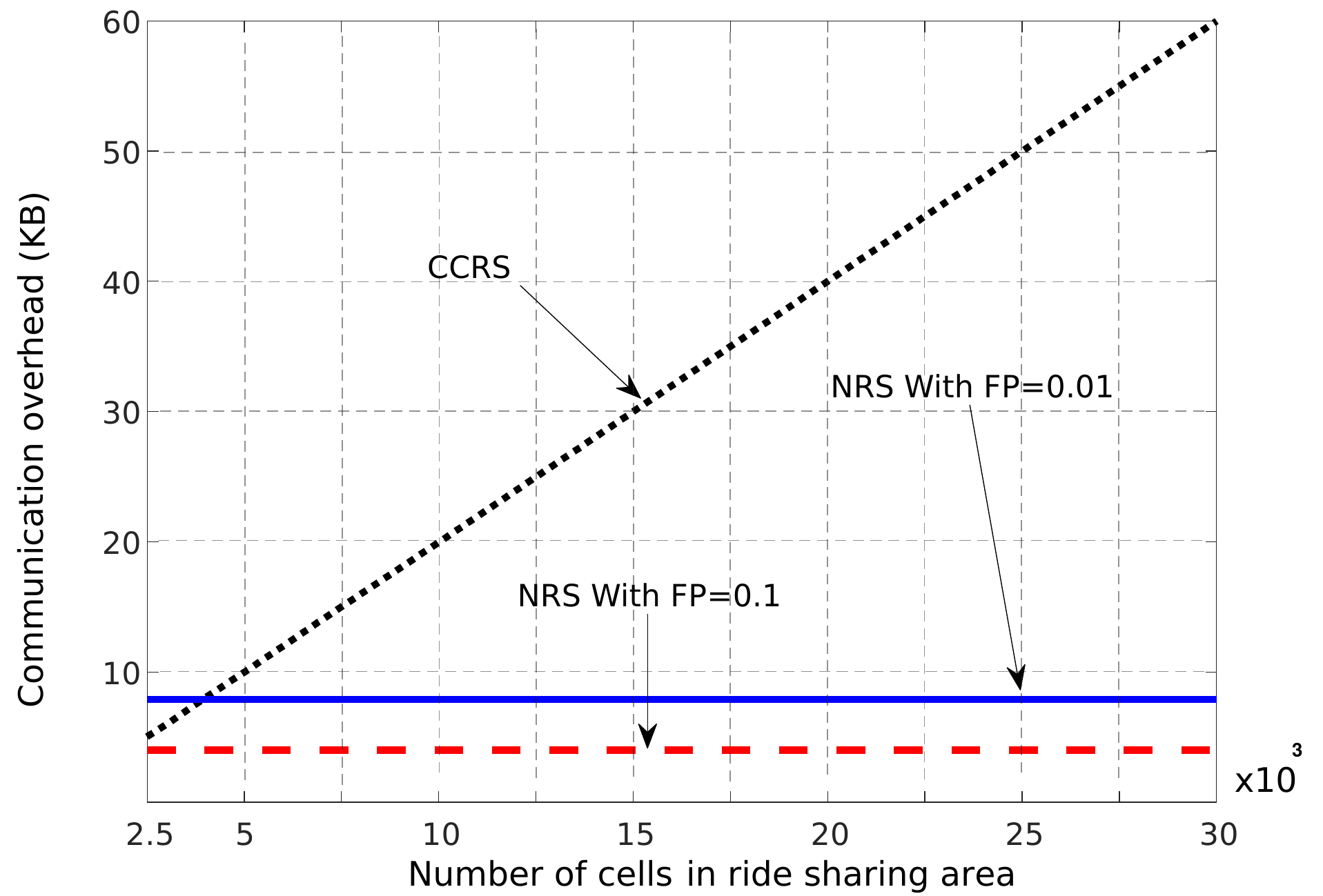}
		\vspace{4mm}
		\caption{Communication overhead versus number of cells in the city.}
		\label{fig:fig13}
	\end{figure}
	

	\textbf{Communication Overhead.} \autoref{fig:fig13} gives the
	communication overhead in KB versus the number of cells in the ridesharing area.
	It compares CCRS to NRS with 0.1 and 0.01 false positive
	probabilities.
	As shown in the figure, for small cities with around 3,000 cells, both schemes have close communication overhead. However, the communication overhead of CCRS increases linearly with the city size while our scheme
	maintains almost a fixed overhead. This can be attributed to the fact
	that in CCRS all the city cells are represented in binary vector before
	encryption, but in our scheme only the pick-up,
	drop-off and route cells are stored in the Bloom filter. Note that, the number of cells can be increased by either increasing the city size or using finer grained cells for the same city for more accurate matching. \autoref{fig:fig13} also illustrates that less false positive rate requires more communication overhead because the size of the Bloom filter should increase, but, the communication overhead required of NRS is still much less than that of CCRS.
	\begin{figure}[t]
		\centering \setlength{\abovecaptionskip}{0mm} \setlength{\belowcaptionskip}{-2mm}
		\includegraphics[clip,scale=0.37]{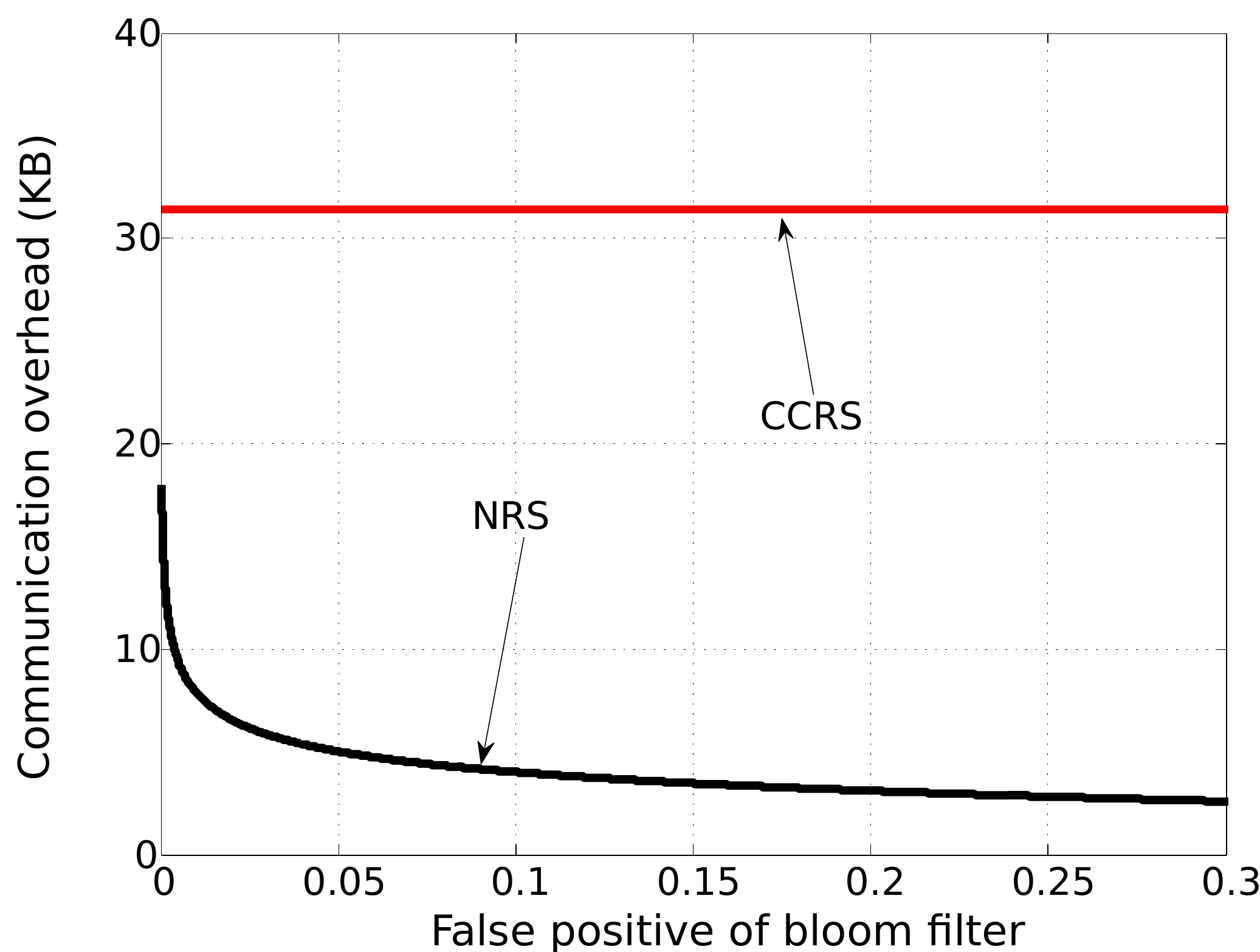}
		\vspace{4mm}
		\caption{Communication overhead versus false positive probability.}
		\label{fig:fig12}

	\end{figure}
	\autoref{fig:fig12} gives the communication overhead for both CCRS and NRS
	versus the false positive probability of the Bloom filter. As shown
	in the figure, CCRS has a constant communication overhead since it
	is free of the false positive probability of the Bloom filter. However for NRS, as the Bloom filter size decreases (lower communication overhead), the false positive probability increases.
	
	Although the false positive probability can be significantly reduced with acceptable communication overhead, the reliability of NRS can be increased more by returning two or three results to reduce the chance of failure. Note that, the failure probability can be reduced exponentially, e.g., $0.1^{R}$ and $0.01^{R}$ in case of $0.1$ and $0.01$ false positive probabilities, where $R$ is
	the number of returned riders.

	\subsection{Transferable ridesharing organization}
	
	\label{sec:performance2}
	
	\subsubsection{Performance Metrics}
	
	Three performance metrics are used for the evaluation of TRS.
	\begin{enumerate}
		\item \textit{Vehicle service rate (SR).} The average number of different requests that each driver can serve.
		\vspace{1mm}
		\item \textit{Ridesharing success rate.} The percentage of
		served rider requests to the total number of requests.
		\vspace{1mm}
		\item \textit{User's preference success rate.} Percentage of satisfied requests based on the type of search preferences specified by users.
	\end{enumerate}

	\subsubsection{Experimental Results and Discussion}

	Table \ref{tab:Computation-Overhead} summarizes the computation overhead
	of our scheme. The required time to generate the secret key for each
	user to be used in the modified kNN encryption scheme is 39.80 ms. It is also
	worth mentioning that the keys can be used for a long period of time.
	For encryption, each driver takes on average 10.30 ms to compute
	$I_{D}^{(r)}$. On the other hand, the rider
	takes on average 0.06 ms to encrypt a request. This time is shorter than route index generation
	because the rider generates only the encryption of the pick-up
	and drop-off location. For the TOS, the graph building of 30 offers
	takes 2.33 sec and Dijkstra's search for one request takes 0.29
	sec.
	\begin{table}[t]
		\center\caption{Computation Overhead. \label{tab:Computation-Overhead}}
		\begin{tabular}{ccc}
		\toprule
		 & \textbf{Operation}  & \textbf{Time}\tabularnewline
		\midrule
		\midrule
		 & User Key Generation  & 39.80 ms\tabularnewline
		\cmidrule{2-3}
		 & Request Index Generation  & 0.06 ms \tabularnewline
		\cmidrule{2-3}
		\textbf{Modified kNN}  & Route Index Generation  & 10.30 ms\tabularnewline
		\cmidrule{2-3}
		\textbf{Encryption Scheme}  & Building Graph with 30 offers  & 2.33 sec\tabularnewline
		\cmidrule{2-3}
		 & Dijkstra's Search for one request  & 0.29 sec\tabularnewline
		\bottomrule
		\end{tabular}
	\end{table}

\begin{figure}[t]
		\vspace{2mm}
		\center\includegraphics[height=6cm, width=8cm]{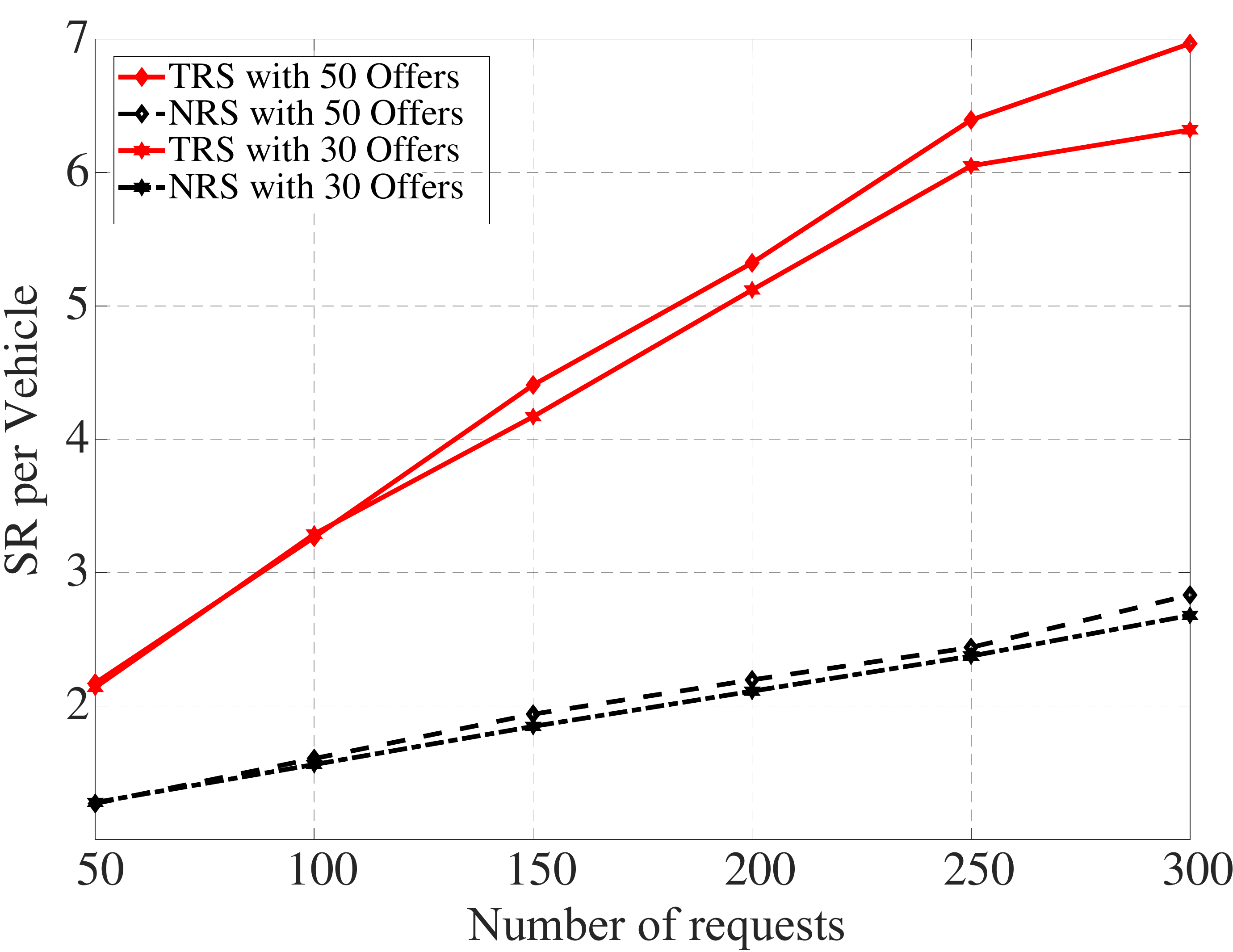} \caption{Vehicle service rate (SR) versus the number of requests for NRS and TRS given different
		numbers of offers. \label{fig:Service-rate-per1}}
	\end{figure}

	\autoref{fig:Service-rate-per1} shows the average SR versus the
	number of requests, with a different number of offers (30 and 50), and studying two rider's preferences (with and without transfer). We assume
	that the maximum number of users that can share a ride in the same
	vehicle is 5. The resolution of the bits of the trip time is set to 25
	bits. The searching preference used in this figure is the $Min_{c}$.
	It can be seen that the SR of the TRS is about double that
	of the NRS. This indicates that more users can be served
	in transferable services. In addition, it shows that the SR
	increases as the number of requests increases. Also, as the number
	of requests increases, each vehicle can serve more than its actual
	capacity as many riders share a ride for only a part of the
	driver's route. However, in NRS, the situation is different
	since the riders' requests should match only one driver's trip data, the SR is decreased. This assure the idea that transferable service can improve each vehicle occupancy, hence, improves the ridesharing utilization.
	
\begin{figure}[!t]
		\center\includegraphics[height=6cm, width=8cm]{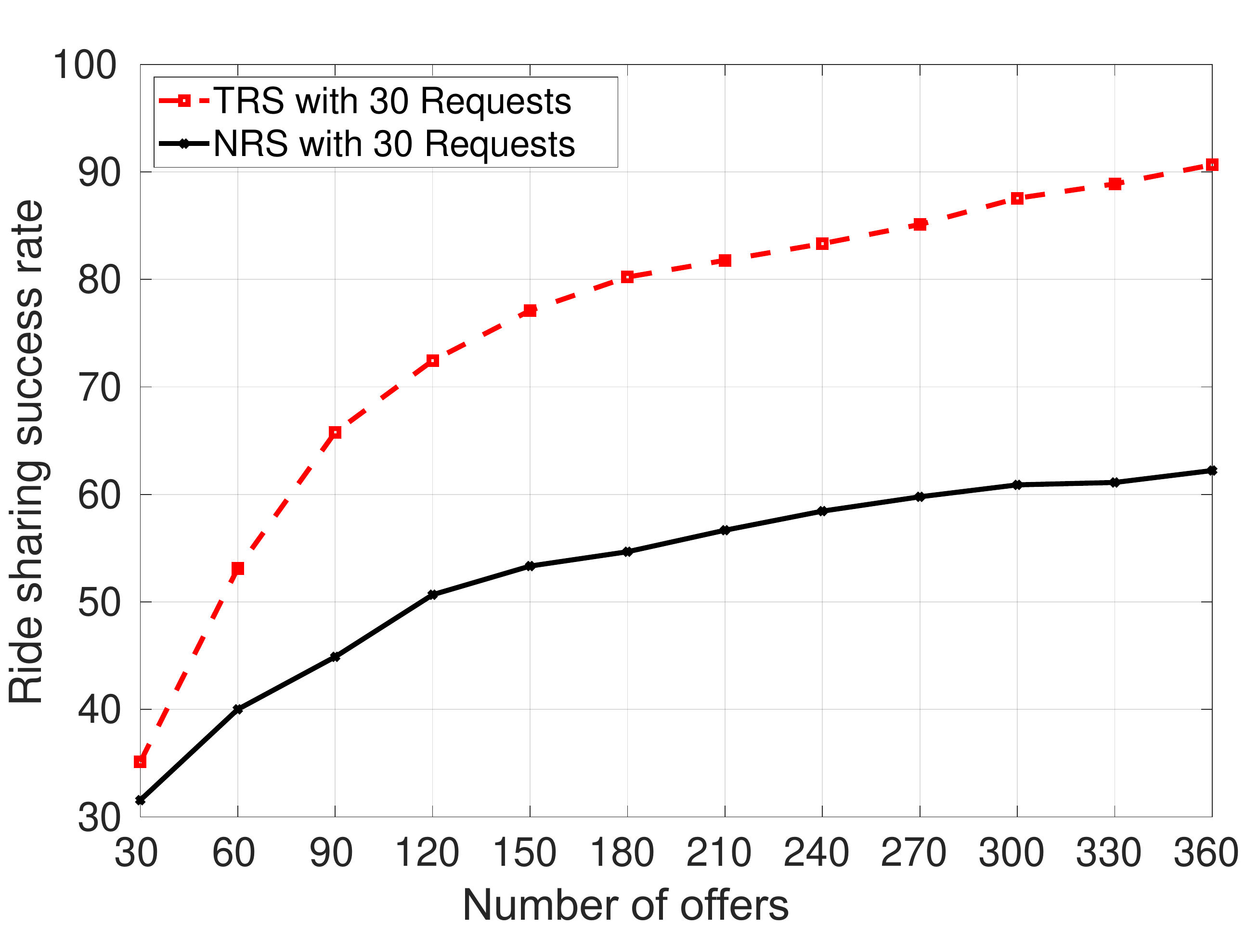}
		
		\caption{Ridesharing success rate for 30 requests as the number of offers increases. \label{fig:Service-Rate-for}}
	\end{figure}

	\autoref{fig:Service-Rate-for} shows the ridesharing success rate versus the
	number of offers in case of 30 requests. The
	maximum number of users in each vehicle is 5 users. The figure shows
	that the ride sharing success rate increases with the increase of the number of offers.
	Also, it is shown that if the number of offers is 330, the ride sharing success rate is 91\% for TRS while it is only 62\% for  NRS. This also proves that transferable service can improve ridesharing success rate, and hence, improves the ridesharing utilization.
	\begin{figure}[t]
		\center\includegraphics[height=6cm, width=8cm]{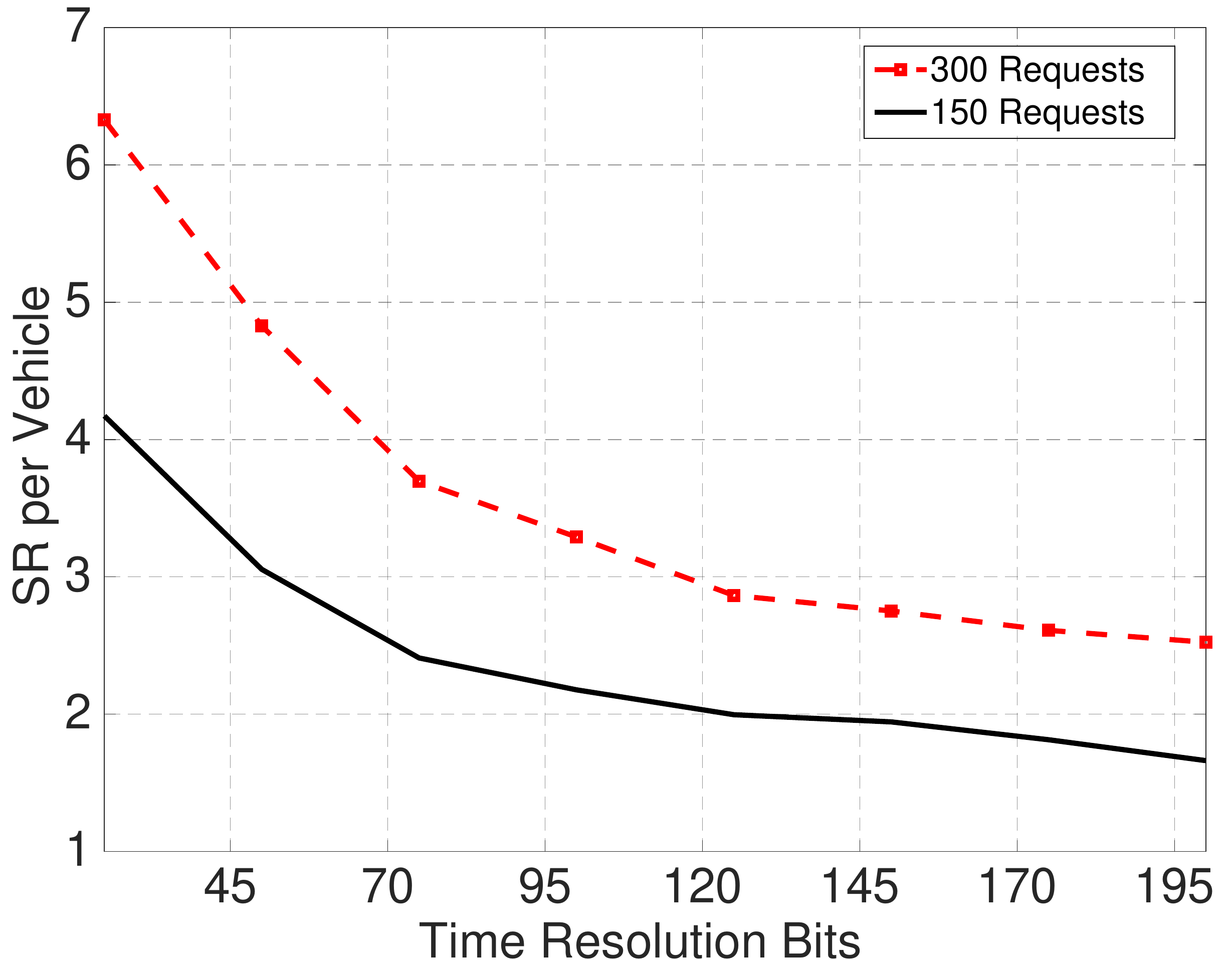} \caption{Effect of the time resolution on the service rate
		per AV. \label{fig:Effect-of-the}} \vspace{-5mm}
	\end{figure}
	
	\begin{figure}[t]
		\vspace{1mm}
		 \center\includegraphics[height=7.5cm, width=9cm]{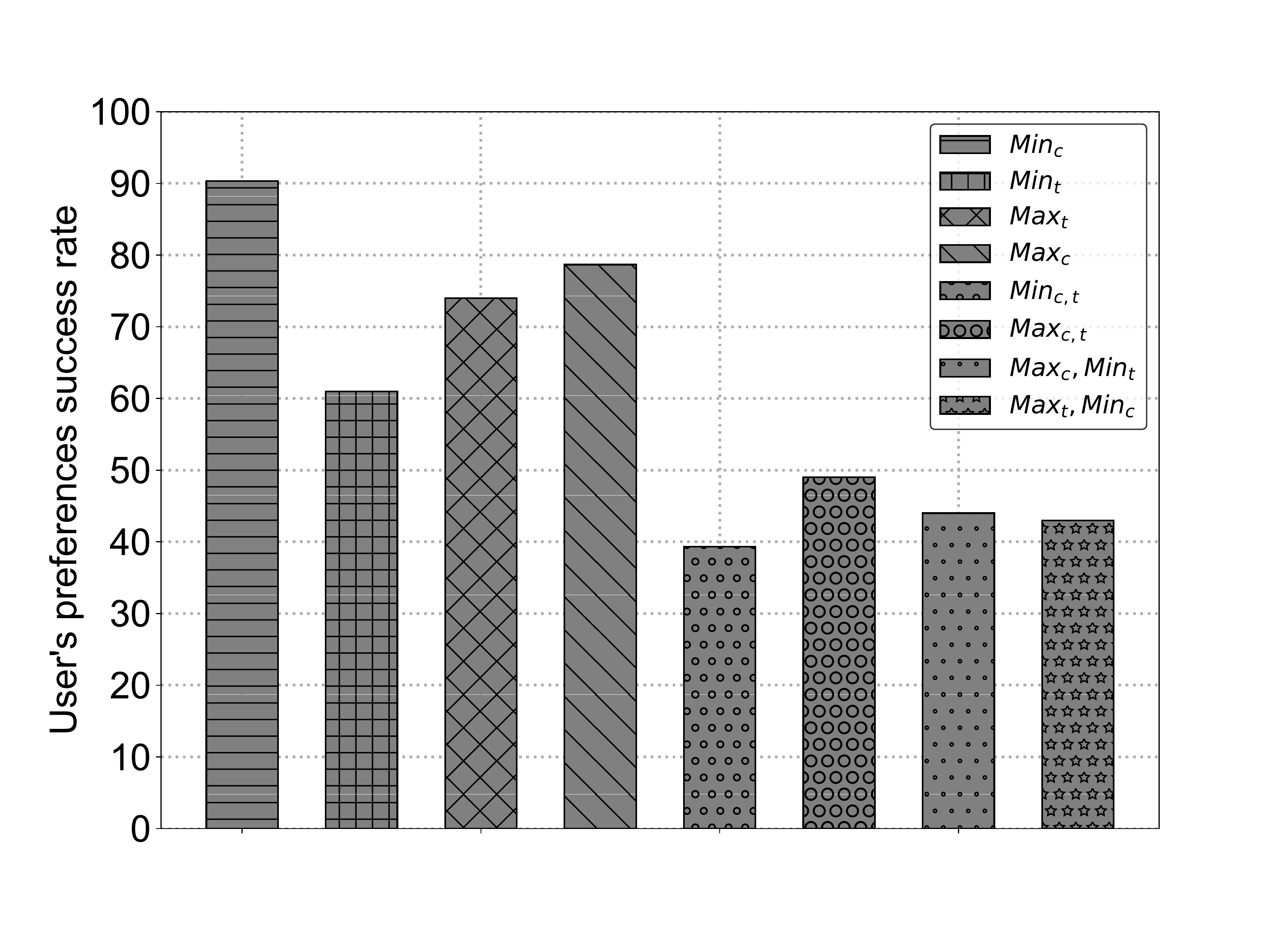} \caption{Success rate for different searching preferences with $T_{c}$ = 30
		and $T_{t}$ = 1. \label{fig:Success-rate-for}} \vspace{-5mm}
	\end{figure}
	
	\autoref{fig:Effect-of-the} gives the effect of the time resolution (in number of bits) on the SR of the vehicle. In this figure, we use TRS with $Min_{c}$ preference and different number of requests (150 and 300).
	As shown in the figure, it can be seen that the increase of the number of time bits decreases the SR of the vehicle. This is because with more time bits, i.e, high resolution, each bit is associated to a shorter period, which decreases the chance of matching offers and requests.
	
	\autoref{fig:Success-rate-for} gives the success rate of the different search preferences used in our scheme. In this figure, we used 30 random requests for each search preference. It can be seen that adding more constrains to the search preference, e.g., cases $Min_{c,t}$, $Max_{c,t}$, and $(Min_c, Max_t)$, reduces the ridesharing success rate, whereas search with a flexible transferable options, has more overall ridesharing success rate.
	\subsection{Comparison between NRS and TRS}
	
	\begin{figure}[!t]
		\centering \includegraphics[clip,height=6cm, width=8.5cm]{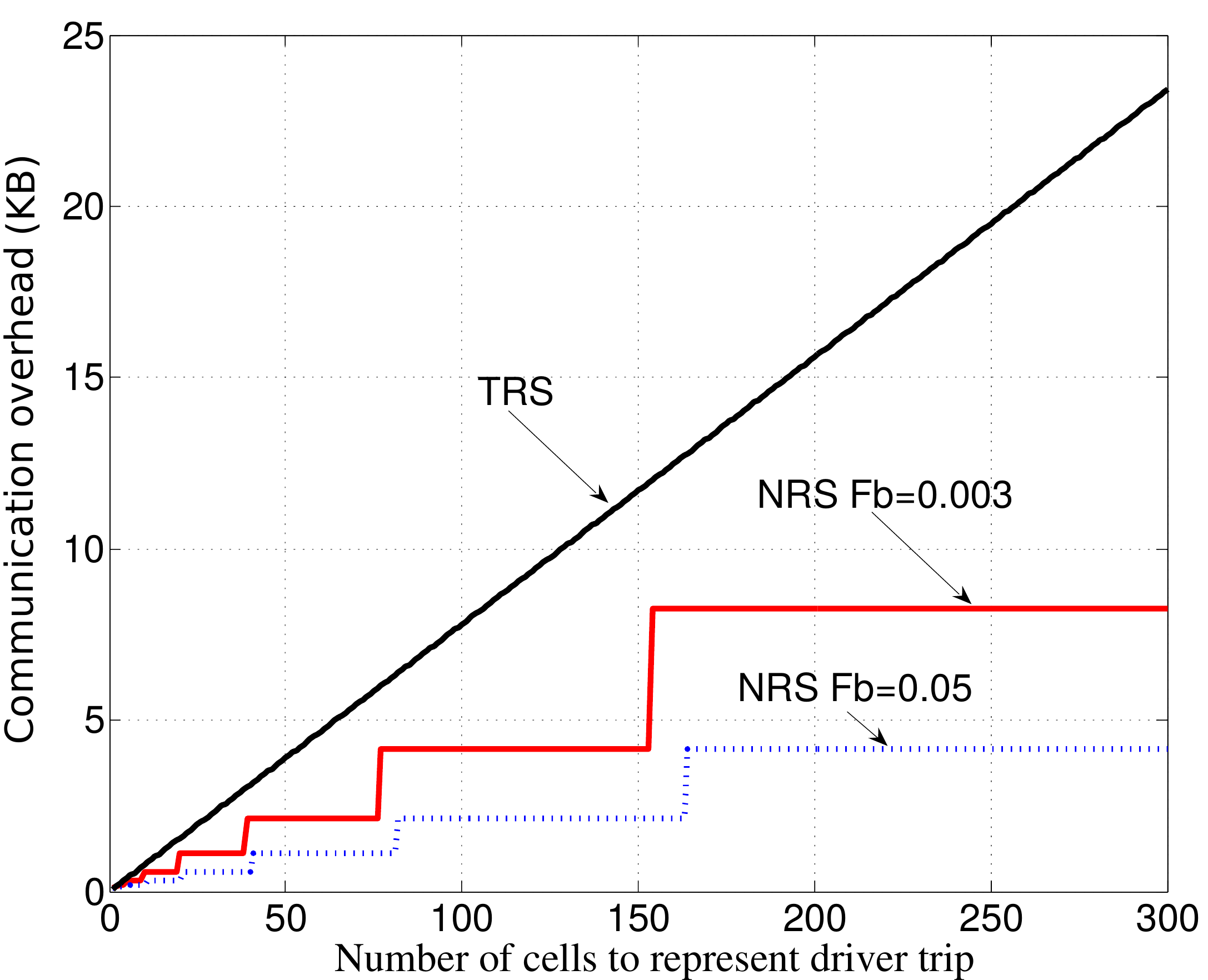}\caption{The communication overhead of NRS and TRS.\label{fig:NRS-vs-TRS}}
		\vspace{-3mm}
	\end{figure}
	
	\autoref{fig:NRS-vs-TRS}
	gives the communication overhead of NRS versus TRS using different
	number of cells to represent the trip data of the driver. For
	NRS, the driver needs to send vectors for pick-up, drop-off,
	and route which
	are represented by Bloom filters with false positive
	probabilities less than $0.003$ and $0.05$ . For TRS, the driver
	needs to send a vector of size $2k+\text{\ensuremath{\ell}}$ bits
	for each cell on its route. From the figure, as the number of cells in the driver route increases, TRS  exhibits a linear communication overhead, because the driver has to send the individual cells in its route for the TOS to build the directed graph. On the other hand, NRS exhibits a ladder steps behavior, due to using Bloom filters where as more cells are added to the filter as the number of cells increases, and thus the filter size should increase in a non-linear behavior to maintain the same false positive probability.
	 From \autoref{fig:NRS-vs-TRS}, although TRS scheme can organize non-transferable shared rides like NRS,
	it has greater communication overhead than NRS. Therfore, it is more efficent to use NRS if users need non-transferable ridesharing service. The trade-off between the communication overhead and the false positive probability for NRS is shown in the figure.
	\begin{figure*}[t]
		\centering
		\subfloat[Time cost in riders' requests]{\label{fig:l1}\includegraphics[width=0.3\textwidth]{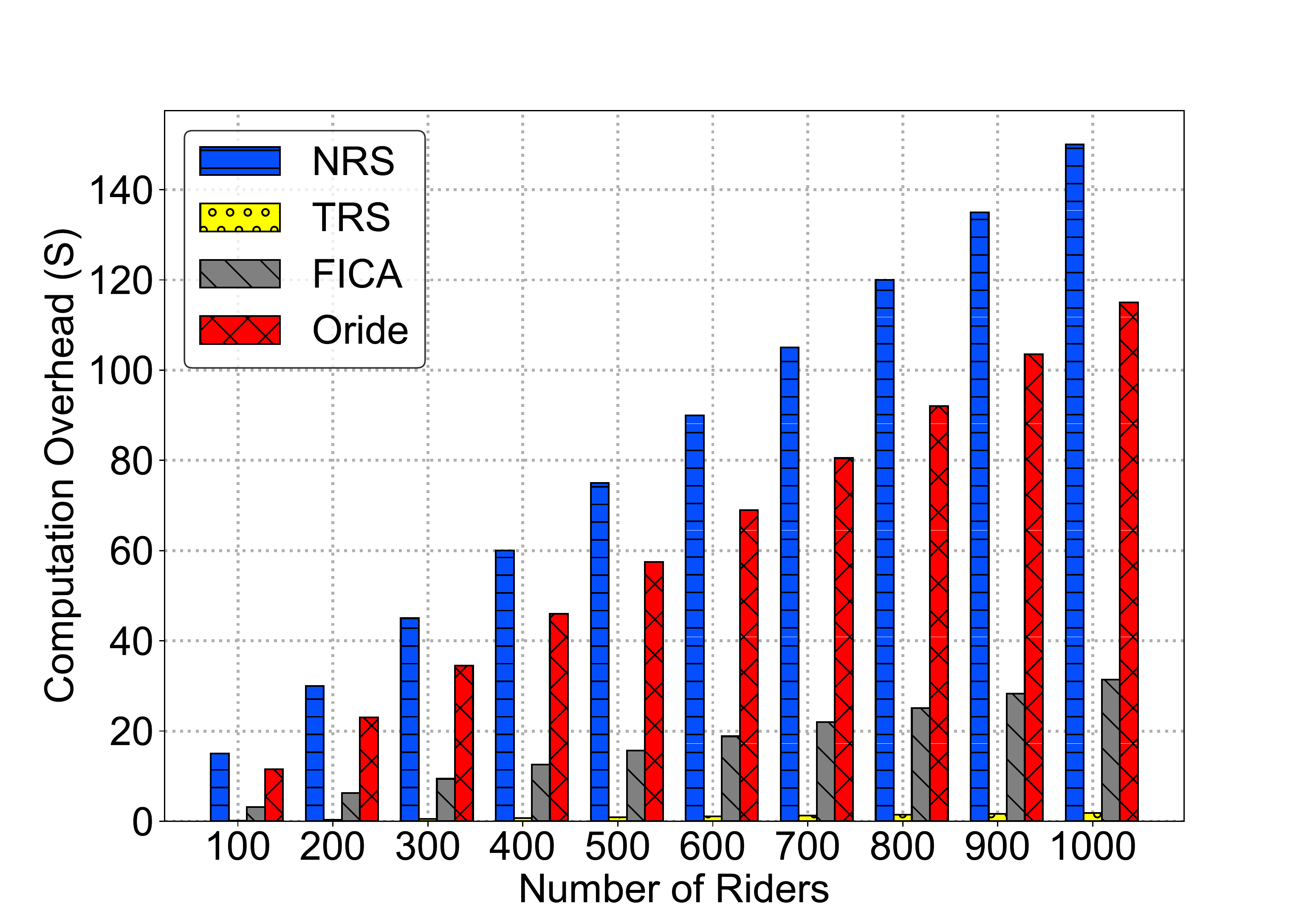}}
		\subfloat[Time cost in drivers' offers]{\label{fig:n1}\includegraphics[width=0.3\textwidth]{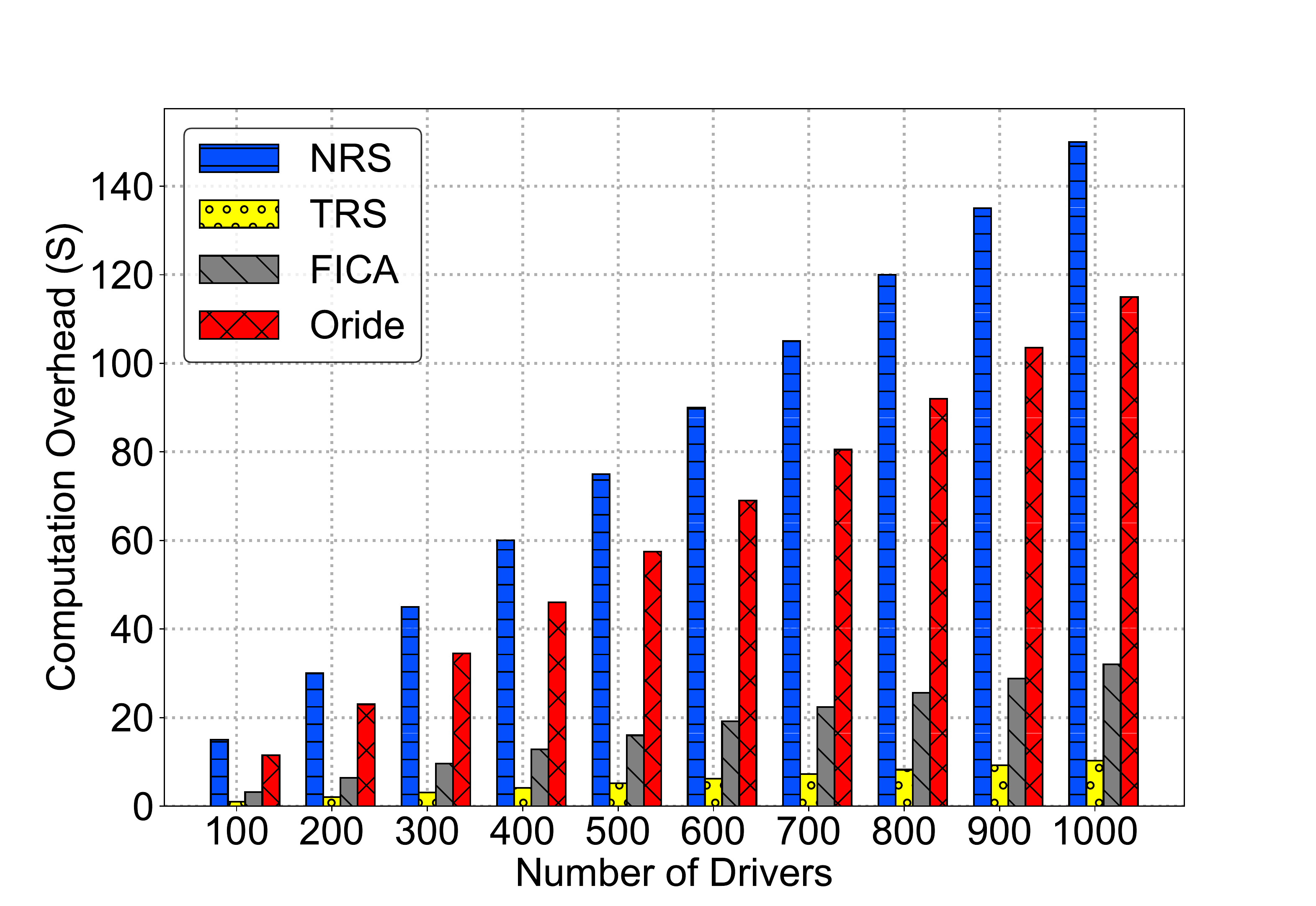}}
		\subfloat[Time cost in ride matching]{\label{fig:o1}\includegraphics[width=0.3\textwidth]{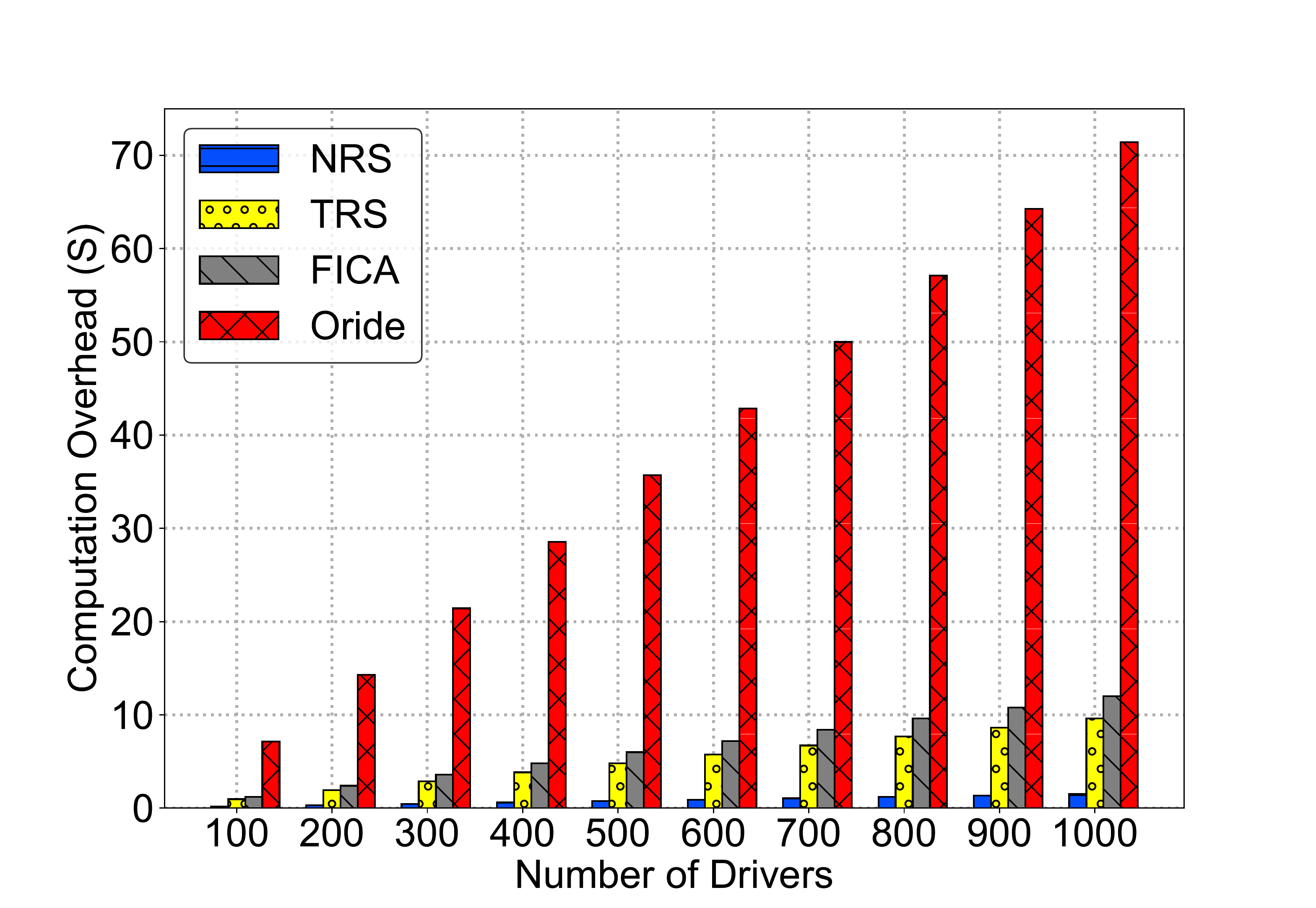}}
		\\
	\subfloat[Communication overhead for riders]{\label{fig:ah1}\includegraphics[width=0.36\textwidth]{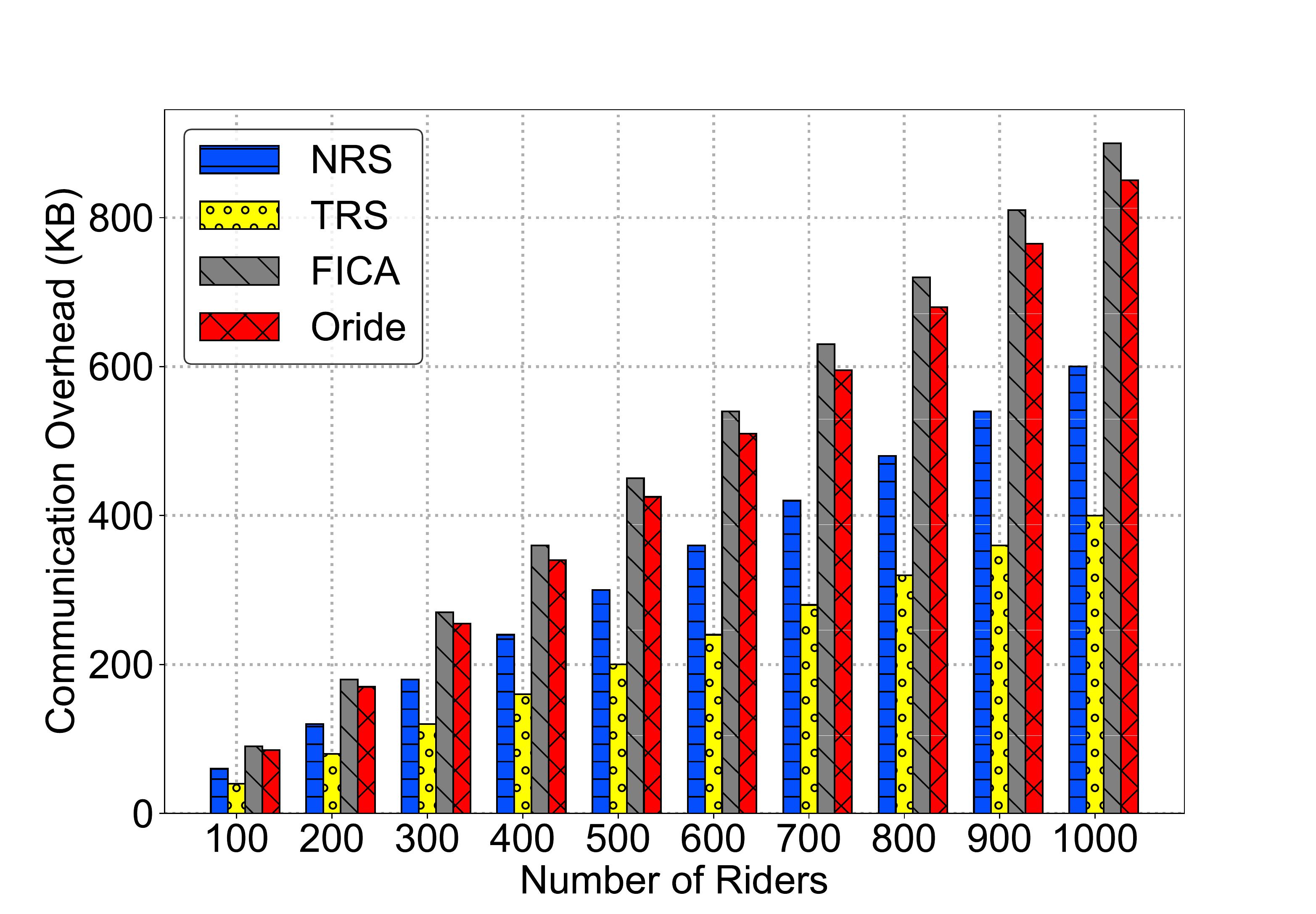}}
	 \ \ \  \subfloat[Communication overhead for drivers]{\label{fig:ao1}\includegraphics[width=0.36\textwidth]{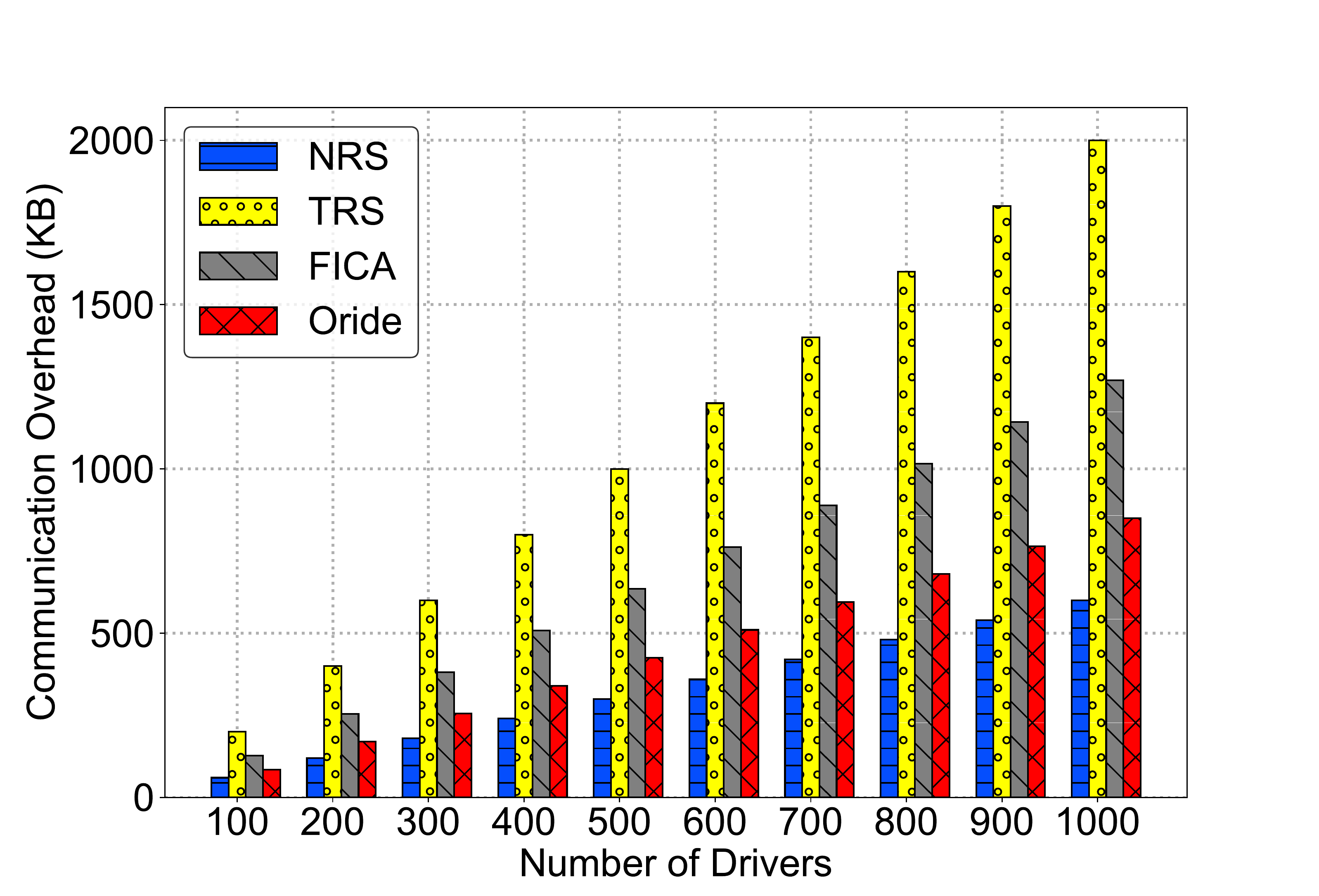}}
		\caption{Performance analysis.}
		\label{fig:GS1}
	\end{figure*}

	\subsection{Comparison with existing schemes} \label{comparison}
We compare the performance of our schemes, NRS and TRS, with existing schemes: FICA \cite {8452961} and ORide \cite {oride2017}.
The following metrics are used for the compution overhead comparison: (1) computation time to generate riders' requests; (2) computation time to generate drivers' offers; and (3) computation time for the ride sharing matching. 
For comparing the communication cost, the following metrics are used: (1) communication overhead for riders' requests; and (2) communication overhead for drivers' offers.

From Figures \ref {fig:l1}, \ref{fig:n1} and \ref{fig:o1}, we can see that our scheme TRS outperforms other schemes in riders' ridesharing request, drivers' ridesharing offers and ride matching computation overhead.
In addition, our TRS scheme offers a transferable service which is not offered in FICA or Oride.
Although, NRS requires more time to compute the rider's request or driver's offer compared with other schemes, it gives the lowest computation overhead in ridesharing matching operation.
As shown in Figure \ref{fig:o1}, our schemes outperform other existing schemes in the computation overhead required for matching the riders' requests with the drivers' offers.

From Figure \ref{fig:ah1}, we can see that our schemes NRS and TRS outperform other schemes in the communication overhead for sending the rider's request.
For NRS, each rider sends only the encrypted bloom filter which is shorter than the actual vector that stores the whole cells in the ridesharing area.
For TRS, each rider only sends his/her pick-up and drop-off locations to request a shared ride.
From Figure \ref{fig:ao1}, we can see that our scheme NRS outperforms other schemes in communication overhead of the driver's offer.
However, TRS requires more communication overhead due to sending the whole route cell by cell to the server to generate the directed graph and enable our unique transferable service.

	\section{Related Work} \label{sec:related Work}
	Recently, research in secure and privacy-preserving  schemes for ride-hailing \cite{Pham2017,Khazbak,oride2017,Wang2018} and ridesharing (carpooling) services \cite{csfHallgren,He2018,ri22,goel2017optimal,Bilogrevic2014,Aivodji2018,Wang2018} is gaining much attention.
Ride-hailing services are motivated by offering high-quality transportation means with affordable cost  (e.g., Uber and Lyft).
It aims to match the riding requests offered by a set of riders to a nearby driver in a real-time fashion through a TOS.
On the other hand, ridesharing services are motivated by the expected reduction in traffic congestion, pollution, and cost when different car owners offer to share their trips with other passengers.
	
Oblivious privacy-preserving ride-hailing services were introduced in \cite{Pham2017,oride2017}.
A scheme, called PrivateRide, is proposed in \cite{Pham2017} to provide anonymous ride-hailing services.
In \cite{oride2017}, a scheme called Oride is proposed to improve PrivateRide by adding accountability and enhanced privacy by increasing the anonymity set for the number of rides from the same area on the same day.
In addition, Oride enables the TOS to revoke any misbehaving riders or drivers.
However, in both PrivateRide and Oride schemes, the TOS has to send to the rider the Euclidean distances for all the drivers so that he can pick the one with the minimum distance which incurs a high communication overhead.
Wang et al. \cite{Wang2018} proposed a privacy-preserving ride-hailing scheme that divides the ridesharing area recursively into quad regions stored into a quadtree.
However, each user needs to send an encryption for each region in the map to report his/her location, which incurs a high communication overhead.

In summary, ride-hailing solutions only aim to connect a rider to the closest driver which is insufficient for ridesharing service due to the fixed route associated with each driver in the latter service.
Besides, unlike previous techniques, in this paper, we used light cryptography to preserve location privacy where the location's cell can be fine-grained with acceptable overhead.
	
For ridesharing services, meeting points determination has been investigated in \cite{Bilogrevic2014} and \cite{goel2017optimal}.
In these schemes, the drivers distributed over a given city using the k-anonymity model aiming to provide good coverage while preserving the riders' privacy.
Nevertheless, the fixed pick-up points may require the driver to take a detour from his intended path while picking up or dropping off riders.
Moreover, as the level of anonymity increases in these schemes, it becomes more challenging to ensure coverage.
Optimal assignment for drivers and riders pairs based on global system parameters has been addressed in \cite{He2018} and \cite{Aivodji2018}.
Ridesharing matching based on secure multi-party computation has been addressed in \cite{Aivodji2018} and \cite{csfHallgren}.
However, all these schemes incur high computation overhead and can not be used to organize transferable rides.
	
The closest work in the literature to this paper is the schemes proposed in \cite{ri22} and \cite {8452961}.
In \cite{ri22} and \cite {8452961}, they address only non-transferable ridesharing service
In \cite{ri22}, a scheme called CCRS has been proposed, where users (both riders and drivers) obscure their trips' data from the TOS by using kNN encryption scheme.
The TOS matches the ciphertext of the users and connects the users that can share rides.
In \cite {8452961}, a scheme called FICA has been proposed.
FICA is secure under a threat model where the cloud server and road side units (RSUs) are semi-honest.
FICA uses an anonymous authentication scheme to authenticate users and recover malicious users' real identities.
In addition, the proposed scheme built a private blockchain into the carpooling system to record carpooling records for data auditability.

Unlike CCRS and FICA, this paper proposes two schemes for non-transferable and transferable services.
The proposed schemes represent the user's route data more efficiently than CCRS, where CCRS represents the whole city cells inside the route vector, setting only the user's route cells to one.
However, in the proposed schemes, every cell has an identifier and the user's route vector encompasses only these identifiers, hence, efficient matching operations.
Moreover, in CCRS, all the riders share the same encryption key, which is less secure.
In our schemes, each user has its own key for encryption.
Different from existing schemes, this work can either be applied to carpooling services or ride-hailing services.
In addition, the proposed schemes offer flexible services selections to both riders and drivers, which requires the TOS to perform several operations over encrypted data.
In particular, we leverage Dijkstra shortest path algorithm to offer transferable ridesharing service, where a rider can transfer between drivers during his/her trip.
This can increase ridesharing utilization without degrading users' privacy.
	
	\section{Conclusion} \label{sec:conclusion}
	In this paper, we proposed an efficient privacy-preserving ridesharing organization schemes for transferable and non-transferable ridesharing services. In NRS, drivers' trip data and riders' requests are compactly stored in Bloom filters which are then encrypted using the modified kNN encryption scheme. Then, TOS can perform similarity measurement over encrypted data to connect each rider with only one driver. In TRS, individual cells of each driver's trip data are encrypted using modified kNN encryption scheme, and the ciphertexts are sent to the TOS. The TOS uses the drivers' trip data to create an encrypted directed graph for ridesharing organization. Riders' preferences are used to determine the weights of the graph's edges. Then, a modified Dijkstra scheme is used by the TOS to organize shared rides by searching the directed graph.
	Our privacy analysis demonstrates that the proposed schemes can preserve users location privacy, trips data privacy, and identity anonymity. Our experimental results on a real map demonstrate
	that the proposed schemes are efficient and can be used to organize shared rides in case of large cities.  Moreover, the results indicate that NRS requires less communication overhead than the TRS. Nevertheless, TRS offers a flexible service that can increase ridesharing utilization, whereas, NRS provides an efficient and useful service for elderly and disabled people who do not prefer to transfer between different drivers.

	\section{Acknowledgement}
	This publication was made possible by NSF grant number CNS-1618549 from the US National Science Foundation. The statements made herein are solely the responsibility of the authors.

	\bibliographystyle{IEEEtran}
	\bibliography{references}		
	\begin{IEEEbiography}[{\includegraphics[width=1in,height=1.25in,clip,keepaspectratio]{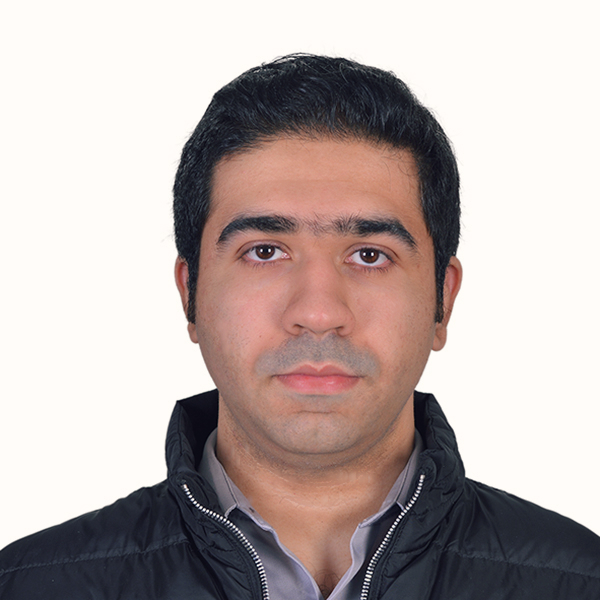}}]{Mahmoud Nabil}
		is currently a Graduate Research Assistant in the Department of Electrical \& Computer Engineering, Tennessee Tech. University, USA and pursuing his Ph.D. degree in the same department.
		He received the B.S. degree and the M.S. degree in Computer Engineering from Cairo University, Cairo, Egypt in 2012 and 2016, respectively.
		His research interests include machine learning, cryptography and network security, smart-grid and AMI networks, and vehicular ad-hoc networks.
	\end{IEEEbiography}

\begin{IEEEbiography}
		[{\includegraphics[width=1in,height=1.25in,clip,keepaspectratio]{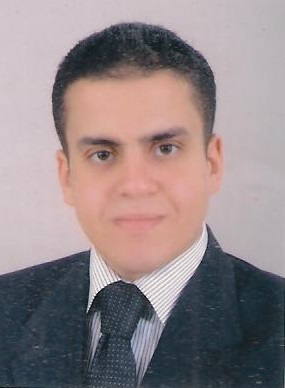}}]{Dr. Ahmed Sherif}
		is an Assistant Professor in the School of Computing Sciences and Computer Engineering at the University of Southern Mississippi (USM). He received his Ph.D. degree in Electrical and Computer Engineering from Tennessee Tech University, Cookeville, Tennessee, United States in August 2017. He received his M.Sc. degree in Computer Science and Engineering from Egypt-Japan University of Science and Technology (E-JUST) in 2014. He joined the Electrical and Computer Engineering department in Tennessee State University (TSU) for one year as a temporary faculty member after finishing his Ph.D. His research interests include security and privacy-preserving schemes in Autonomous Vehicles (AVs), Vehicular Ad hoc Networks (VANETs), Internet of Things (IoT) applications, and Smart Grid Advanced Metering Infrastructure (AMI) network.
		\end{IEEEbiography}
	
	\begin{IEEEbiography}[{\includegraphics[width=1in,height=1.25in,clip,keepaspectratio]{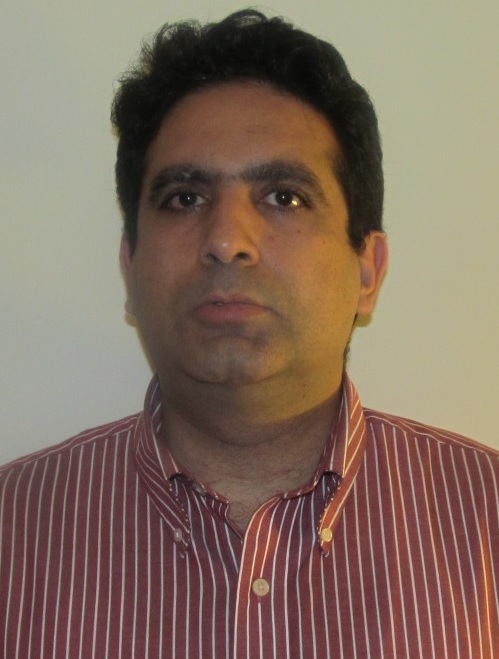}}]{Dr. Mohamed M. E. A. Mahmoud}
		received PhD degree from the University of Waterloo in April 2011. From May 2011 to May 2012, he worked as a postdoctoral fellow in the Broadband Communications Research group - University of Waterloo. From August 2012 to July 2013, he worked as a visiting scholar in University of Waterloo, and a postdoctoral fellow in Ryerson University. Currently, Dr Mahmoud is an associate professor in Department Electrical and Computer Engineering, Tennessee Tech University, USA. The research interests of Dr. Mahmoud include security and privacy preserving schemes for smart grid communication network, mobile ad hoc network, sensor network, and delay-tolerant network. Dr. Mahmoud has received NSERC-PDF award. He won the Best Paper Award from IEEE International Conference on Communications (ICC'09), Dresden, Germany, 2009. Dr. Mahmoud is the author for more than twenty three papers published in major IEEE conferences and journals, such as INFOCOM conference and IEEE Transactions on Vehicular Technology, Mobile Computing, and Parallel and Distributed Systems. He serves as an Associate Editor in Springer journal of peer-to-peer networking and applications. He served as a technical program committee member for several IEEE conferences and as a reviewer for several journals and conferences such as IEEE Transactions on Vehicular Technology, IEEE Transactions on Parallel and Distributed Systems, and the journal of Peer-to-Peer Networking.
	\end{IEEEbiography}
	
	\begin{IEEEbiography}[{\includegraphics[width=1in,height=1.25in,clip,keepaspectratio]{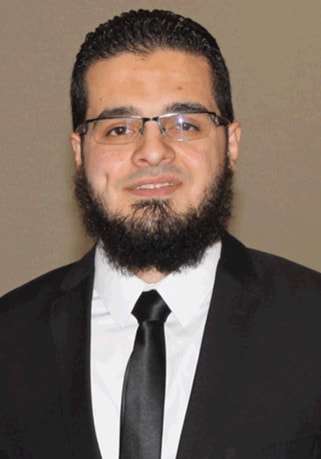}}]{\textbf{Dr. Ahmad Alsharif}}
		(M'18) received the Ph.D. in Electrical and Computer Engineering from Tennessee Tech. University in 2019.
		He received the B.Sc. and M.Sc. degrees in Electrical Engineering from Benha University, Egypt in 2009 and 2015, respectively.
		In 2009, he was one of the recipients of the young innovator award from the
		Egyptian Industrial Modernisation Centre.
		Currently, Dr. Alsharif is an assistant professor of cybersecurity at the University of Central Arkansas. His research interests include security and privacy in smart grid, vehicular Ad Hoc networks, multihop cellular wireless networks, and cyber-physical systems.
	\end{IEEEbiography}
		
	\begin{IEEEbiography}
		[{\includegraphics[width=1in,height=1.25in,clip,keepaspectratio]{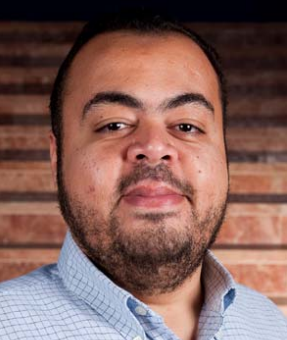}}]{Dr. Mohamed Abdallah}
		(S'94-M'08-SM'13) was born in Giza, Egypt. He received the B.Sc. degree with honors from Cairo University, Giza, Egypt, in 1996, and the M.Sc. and Ph.D. degrees in electrical engineering from University of Maryland at College Park, College Park, MD, USA, in 2001 and 2006, respectively. He joined Cairo University in 2006 where he holds the position of Associate Professor in the Electronics and Electrical Communication Department. He is currently an Associate Research Scientist at Texas A\&M University at Qatar, Doha, Qatar. His current research interests include the design and performance of physical layer algorithms for cognitive networks, cellular heterogeneous networks, sensor networks, smart grids, visible light and  free-space optical communication systems and reconfigurable smart antenna systems.
		\end{IEEEbiography}

\end{document}